\documentclass{elsart}

\usepackage{amssymb}

\newcommand{\Boost}{\hat{B}_c}

\newcommand{\ophi}{\hat{\phi}}
\newcommand{\opi}{\hat{\pi}}

\newcommand{\oD}{\hat{D}}
\newcommand{\oa}{\hat{a}}
\newcommand{\oadag}{\hat{a}^\dag}
\newcommand{\ob}{\hat{b}}
\newcommand{\obdag}{\hat{b}^\dag}

\begin{document}

\begin{frontmatter}



\title{Point-form quantum field theory}

\author[Graz]{E.P. Biernat},
\ead{elmar.biernat@stud.uni-graz.at}
\author[Iowa]{W.H. Klink},
\ead{william-klink@uiowa.edu}
\author[Graz]{W. Schweiger},
\ead{wolfgang.schweiger@uni-graz.at}
\author[Graz]{S. Zelzer\thanksref{now}}
\thanks[now]{Present address: Deutsches Krebsforschungszentrum,
D-69009 Heidelberg, Germany}
\ead{s.zelzer@dkfz-heidelberg.de}
\address[Graz]{Institut f\"ur Physik, Universit\"at Graz, A-8010 Graz, Austria}
\address[Iowa]{Department of Physics and Astronomy, University of Iowa, Iowa City, IA, USA}

\begin{abstract}
We examine canonical quantization of relativistic field theories
on the forward hyperboloid,  a Lorentz-invariant surface of the
form
 $x_\mu x^\mu = \tau^2$.
This choice of quantization surface implies that all components of
the 4-momentum operator are affected by interactions (if present),
whereas rotation and boost generators remain interaction free -- a
feature characteristic of Dirac's \lq\lq point-form\rq\rq of
relativistic dynamics. Unlike previous attempts to quantize fields
on space-time hyperboloids, we keep the usual plane-wave expansion
of the field operators and consider evolution of the system
generated by the 4-momentum operator.  We verify that the
Fock-space representations of the Poincar\'e generators for free
scalar and spin-1/2 fields look the same as for equal-time
quantization. Scattering is  formulated for interacting fields in
a covariant interaction picture and it is shown that the familiar
perturbative expansion of the S-operator is recovered by our
approach. An appendix analyzes
 special distributions, integrals over the forward hyperboloid, that are used repeatedly in the paper.
 \end{abstract}
\end{frontmatter}

\section{Introduction}
\label{intro}
One of the main goals of formulating a relativistic many-body
theory is to find a realization of the Poincar\'e algebra in terms
of operators which act on a Fock space. From knowledge of the
Poincar\'e generators it is then possible to calculate physical
observables like the mass spectrum, scattering data, or
probability distributions in different inertial frames. For
systems of free particles the Fock-space representation of the
Poincar\'e generators follows immediately from their
representations on one-particle Hilbert spaces. For interacting
many-body systems the situation is much more complicated. A close
inspection of the Poincar\'e algebra reveals that interaction
terms must appear in more than one Poincar\'e generator and are,
in general, constrained by non-linear relations which are hard to
satisfy. A possible solution to this problem is to start with a
classical field theory, which is specified by a (Lorentz-)scalar
Lagrangian density, and quantize it. The operators which generate
Poincar\'e transformations then automatically satisfy the
Poincar\'e algebra, even for interacting theories. According to
Tomonaga~\cite{t46} and Schwinger~\cite{s48} (see also
Ref.~\cite{s61}) quantization can be carried out on an arbitrary
space-like hypersurface of Minkowski space-time by imposing
(generalized) canonical (anti)commutation relations on the field
operators. Evolution of the system from one such hypersurface to
the neighboring one may then be described in a Lorentz-invariant
way by means of the, so called, \lq\lq Tomonaga-Schwinger
equation\rq\rq.

We make use of this general framework and quantize field theories
on the Lorentz-invariant space-time hyperboloid $\Sigma:\, x_\mu
x^\mu = \tau^2$, $\tau$ arbitrary but fixed. For interacting
theories the choice of the quantization surface is intimately
connected with the interaction dependence of the Poincar\'e
generators. In our case all components of the 4-momentum operator
become interaction dependent, whereas the generators of Lorentz
transformations stay free of interactions. This resembles Dirac's
point form of classical relativistic dynamics~\cite{d49}.
Therefore we speak of \lq\lq point-form quantum field theory\rq\rq
(PFQFT). For the usual equal-time quantization, which corresponds
to Dirac's instant form, interaction terms appear in the generator
of time translations, i.e. $\hat{P}^0$, and in the generators of
Lorentz boosts $\hat{K}_i$, $i=1,2,3$. In either case, the
interaction-free generators give rise to a subgroup of the
Poincar\'e group, the so called, \lq\lq kinematic subgroup\rq\rq\
or \lq\lq stability group\rq\rq, which leaves the quantization
surface invariant. Since there exist 3 additional continuous
subgroups of the Poincar\'e group one may think of other preferred
choices of quantization surfaces~\cite{ls77}. Actually, only field
quantization at equal time $t$, equal light-cone time $x_+=t+x_3$,
and on space-time hyperboloids $x_\mu x^\mu = \tau^2$ has been
discussed in the literature. Equal-time quantization is common
text-book knowledge. Field quantization on the light front is also
well developed and has attracted interest in connection with hard
hadronic processes and with the solution of the QCD bound-state
problem~\cite{c92,bp98}.

But only a few old papers exist which are dedicated to
PFQFT~\cite{f73,s74,grs74,ds74a,ds74b,note}. The reason is, of
course, the curved nature of the quantization surface which poses
some technical problems. Nevertheless, from a conceptual point of
view PFQFT is rather attractive. The interaction-dependent,
dynamical Poincar\'e generators are components of a 4-vector and
the interaction-free, kinematical Poincar\'e generators can be
combined to a second-order tensor. This makes a manifestly
Lorentz-covariant formulation of PFQFT feasible, as had already
been noticed by Dirac~\cite{d49}.

The benefits of the point form have up till now only been
exploited in relativistic quantum mechanics for the analysis of
electromagnetic current operators~\cite{l95,k03a,m06a} and the
calculation of masses~\cite{g98, g04}, electroweak
properties~\cite{b01,b04} and strong decays of hadrons~\cite{m06b,
m06c} within constituent quark models. Ref.~\cite{k03b} takes
advantage of point form relativistic quantum mechanics to develop
a Poincar\'e invariant coupled-channel formalism in which the
interaction vertices are derived from quantum field theory. First
applications of this formalism include the calculation of (axial)
vector-meson masses and decays within the chiral constituent quark
model~\cite{k03c,k05}. This coupled-channel formalism can also be
understood as a truncation scheme for quantum field theories which
preserves Poincar\'e invariance. The additional approximation
which enters is the assumption that the total velocity of the
system is conserved at interaction vertices. In fact, quantum
field theoretical ideas form the background for most applications
of point form relativistic quantum mechanics. The starting point
for the construction of particle-exchange potentials or current-
and decay-operators which are used in relativistic quantum
mechanics is usually a quantum field theory. It is thus important
 to put the operator formalism for field
quantization on space-time hyperboloids on a solid footing.

Since we aim at a Fock-space representation of Lagrangian field
theories we have to specify a Fock space. Fock spaces are infinite
direct sums of tensor products of single-particle Hilbert spaces.
Each single-particle space is a representation space for a unitary
irreducible representation of the Poincar\'e group. General
one-particle states, representing a particle with certain mass and
spin, may thus be expressed as a superposition of eigenstates of a
complete set of commuting self-adjoint operators that is
constructed from the Poincar\'e generators. The basis most
commonly used is the, so called, \lq\lq Wigner basis\rq\rq\ which
consists of simultaneous eigenstates of the 3-momentum operator
and an additional operator describing the spin orientation. This
basis diagonalizes part of the generators of the kinematic
subgroup for equal-time quantization, i.e. those of the Abelian
subgroup of translations. However, the old papers on PFQFT made
use of another basis, which was obtained by reducing the
Poincar\'e group with respect to the Lorentz subgroup, the, so
called, \lq\lq Lorentz basis\rq\rq. In this basis the Casimir
operator of the homogeneous Lorentz group as well as the operators
for the total angular momentum and one of its components are
diagonalized simultaneously~\cite{grs74}. But the big disadvantage
of such a basis is that the 4-momentum operator cannot easily be
defined as a self-adjoint operator acting on square-integrable
functions~\cite{mr72}. Since a primary goal is to show the
equivalence of quantization on space-time hyperboloids and
equal-time quantization for free and simple interacting field
theories we will stick with the usual basis of eigenstates of the
(free) 3-momentum operator.

Often point-form is associated with evolution of the system in the
parameter $\tau$, i.e. perpendicular to the hyperboloid on which
field quantization takes place. This kind of evolution is
generated by the dilatation operator and has been studied in the
old papers on PFQFT~\cite{f73,s74,grs74,ds74a,ds74b}. At first
sight it seems to be quite natural to consider evolution in $\tau$
as soon as one introduces hyperbolic coordinates to parameterize
the hyperboloid $x_\mu x^\mu = \tau^2$. But evolution in $\tau$
also gives rise to some problems. If one is not dealing with a
massless theory, the dilatation operator is $\tau$-dependent. With
hyperbolic coordinates one needs at leat 3 different coordinate
patches to cover the whole Minkowski space-time~\cite{ds74b}. It
thus becomes rather cumbersome to follow the evolution of the
system from the backward to the forward light cone as is, e.g.,
necessary if one wants to formulate scattering. Looking back at
Dirac's seminal paper on Hamiltonian formulations of classical
relativistic dynamics no reference is made to a particular choice
of a time parameter. The different forms are only characterized by
the space-like hypersurface of Minkowski space-time on which the
initial conditions are posed and those Poincar\'e generators which
do not generate the kinematic subgroup are denoted as \lq\lq
Hamiltonians\rq\rq~\cite{d49}. The Hamiltonians tell us how the
dynamical variables of the system evolve under the corresponding
Poincar\'e transformations. Their knowledge suffices to calculate
the evolution of the system from the distant past to the far
future in any inertial frame. The situation is quite the same for
the operator approach to quantum field theories. In the case of
PFQFT the Fock-space representation of the 4-momentum operator
already contains all the information which is necessary for the
calculation of the mass spectrum and the scattering matrix. It
will thus be a primary task to show how the Fock-space
representation of the 4-momentum operator looks if a spin-zero or
spin-1/2 field is quantized on the forward hyperboloid. Since we
will use the usual momentum state basis, differences with
equal-time quantization are only to be expected for interacting
fields.

Sec.~\ref{history} elucidates the problems encountered in previous
attempts to formulate PFQFT for the simplest case of a real scalar
field theory in 1+1 dimensional space-time. After having realized
that these problems are mainly connected with the Lorentz basis
and the evolution in $\tau$, we will switch to the usual Wigner
basis and concentrate on the evolution of the system generated by
the 4-momentum operator. The equivalence of equal-$\tau$ and
equal-time quantization for free spin-0 and spin-1/2 fields is
proved in Sec.~\ref{quantization}. Thereby it turns out that all
the necessary integrations over the forward hyperboloid can be
carried out in Cartesian coordinates with the help of an
appropriately defined distribution $W(P,Q)$. A manifestly
covariant formulation of scattering is then developed in
Sec.~\ref{scattering}. It is shown that the perturbative expansion
of the S-operator is equivalent to usual time-ordered perturbation
theory. A summary of our findings and an outlook to further
applications can be found in Sec.~\ref{summary}. Finally the
distribution $W(P,Q)$ and its properties are discussed in some
detail in App.~\ref{WPQ}.

\section{Historical attempts}
\label{history}
To the best of our knowledge all the historical papers on PFQFT in
Minkowski space-time did not go far beyond free
fields~\cite{s74,grs74,ds74b}. The reasons can already be
demonstrated for the simplest case of a real scalar field in 1+1
dimensional space-time. To see this we briefly recall the strategy
of Refs.~\cite{s74,ds74b}. The starting point is the Lagrangian
density, which for a free massive scalar field takes on the form
\begin{equation}
\label{eq:L11} \mathcal{L}(t,x)=\frac{1}{2}\left[ (\partial_t
\phi(t,x))^2 - (\partial_x \phi(t,x))^2 - m^2 \phi^2(t,x)\right]\,
.
\end{equation}
Since we want to quantize the theory on the hyperboloid $t^2-x^2 =
\tau^2$ we go over to hyperbolic coordinates (with
$\tau=e^\alpha$)
\begin{equation}
t = e^\alpha \, \cosh \beta \, ,\quad x = e^\alpha \, \sinh \beta
\, , \quad -\infty<\alpha, \beta< \infty \, .
\end{equation}
With this change of coordinates our considerations are restricted
to the forward light cone. The Lagrangian density expressed in
terms of hyperbolic coordinates is
\begin{equation}
\mathcal{L}(\alpha,\beta)=\frac{1}{2}\left[ (\partial_\alpha
\phi(\alpha,\beta))^2 - (\partial_\beta \phi(\alpha,\beta))^2 -
e^{2 \alpha} m^2 \phi^2(\alpha,\beta)\right]\, .
\end{equation}
Via the action principle it gives rise to the Klein-Gordon
equation in hyperbolic coordinates:
\begin{equation}
\label{KG} \left( \partial_\alpha^2 - \partial_\beta^2 + e^{2
\alpha} m^2 \right)\phi(\alpha,\beta) = 0 \, .
\end{equation}
Since $\beta$ parameterizes the hyperboloid it is natural to
consider $\alpha$ as time parameter and proceed analogous to
canonical quantization at equal time (with $x$ replaced by $\beta$
and $t$ by $\alpha$). The canonical momentum conjugate to
$\phi(\alpha, \beta)$ is
\begin{equation}
\pi(\alpha, \beta) = \frac{\partial
\mathcal{L}(\alpha,\beta)}{\partial(\partial_\alpha \phi(\alpha,
\beta))}\, ,
\end{equation}
and the Hamiltonian (i.e. the Legendre transform of the
Lagrangian) which generates \lq\lq translations\rq\rq\ in $\alpha$
may be identified (in 1+1 dimensional space-time) with the
dilatation generator
\begin{equation}
D(\alpha) = \frac{1}{2} \int_{-\infty}^\infty \, \left[
(\partial_\alpha \phi(\alpha,\beta))^2 + (\partial_\beta
\phi(\alpha,\beta))^2 + e^{2 \alpha} m^2
\phi^2(\alpha,\beta)\right]\, d\beta \, .
\end{equation}
Note that $D(\alpha)$ is explicitly $\alpha$-dependent (i.e.
$\partial_\alpha D(\alpha)\neq 0$) and does not belong to the set
of Poincar\'e generators.

According to the rules for canonical quantization the classical
fields $\phi(\alpha, \beta)$ and $\pi(\alpha, \beta)$ have now to
be replaced by corresponding operators $\hat\phi(\alpha, \beta)$
and $\hat\pi(\alpha, \beta)$, respectively, which have to satisfy
equal-$\alpha$ commutation relations
\begin{eqnarray}
\label{ccr}
\left[ \ophi({\alpha},\beta) ,
\opi({\alpha},\beta^\prime) \right] &=& i
\delta(\beta-\beta^\prime)\, , \nonumber \\ \left[ \ophi({
\alpha},\beta) , \ophi({\alpha},\beta^\prime) \right] &=& \left[
\opi({\alpha},\beta) , \opi({\alpha},\beta^\prime) \right] = 0 \,
.
\end{eqnarray}
Here we have used the Heisenberg representation. The Heisenberg
equations of motion
\begin{eqnarray}
\label{heisen}
\partial_\alpha \ophi (\alpha,\beta) &=& - i \left[
\ophi(\alpha,\beta), \oD(\alpha) \right] = \opi (\alpha,\beta)\, ,
\nonumber \\
\partial_\alpha\opi (\alpha,\beta) &=& - i \left[ \opi(\alpha,\beta), \oD(\alpha)
\right]\, ,
\end{eqnarray}
follow immediately from the canonical commutation relations,
Eqs.~(\ref{ccr}), with $\oD(\alpha)$ being the (quantized)
dilatation operator. Equations~(\ref{heisen}) further imply that
the field operator $\ophi(\alpha,\beta)$ satisfies the original
Klein-Gordon equation, Eq.~(\ref{KG}).

The construction of the Fock space usually starts with a choice of
basis states for the one-particle Hilbert space. To this aim the
field operator is expanded in terms of solutions of the
Klein-Gordon equation, Eq.~(\ref{KG}), which are orthogonal under
the $\alpha$-independent scalar product
\begin{equation}
(\phi,\psi)_\alpha = i \int_{-\infty}^{\infty} \, d\beta\,
\left[\phi^\ast (\alpha^\prime , \beta) \frac{\partial
\psi(\alpha^\prime , \beta)}{\partial \alpha^\prime} -
\frac{\partial \phi^\ast (\alpha^\prime , \beta)}{\partial
\alpha^\prime} \psi (\alpha^\prime , \beta)
\right]_{\alpha^\prime=\alpha} \, .
\end{equation}
Imitating equal-time quantization it is assumed that the $\alpha$-
and $\beta$-depen\-den\-ce of these solutions factorizes and the
$\beta$-dependent part is simply a plane wave. An appropriately
normalized ($(\phi_\lambda, \phi_{\lambda^\prime})_\alpha =
\delta(\lambda - \lambda^\prime)$) complete set of solutions,
which meets these requirements, is given by
\begin{eqnarray}
\phi_\lambda (\alpha, \beta) &=& -\frac{i e^{\frac{\pi}{2}
\lambda}}{\sqrt{8}}  {H_{i \lambda}^{(2)}(m e^\alpha)} \, {e^{i
 \lambda \beta}} \qquad \hbox{and} \nonumber\\ \phi_\lambda^\ast(\alpha,\beta)
 &=& \frac{i e^{-\frac{\pi}{2} \lambda}}{\sqrt{8}}  {H_{i \lambda}^{(1)}(m e^\alpha)}
 \, e^{-i \lambda \beta}\, ,
\end{eqnarray}
with $-\infty < \lambda < \infty$ and $H_{i \lambda}^{(\cdot)}$
denoting Hankel functions. The reason for taking Hankel functions
for the $\alpha$-dependence is that they satisfy the same boundary
conditions as in equal-time quantization, i.e.
$\phi_\lambda(\alpha, \beta)$ are solutions of the Klein-Gordon
equation which travel forward in (ordinary) time $t$.
Consequently, these solutions lead to the usual Feynman
propagator~\cite{ds74b}. Ref.~\cite{s74}, on the other hand, has
taken solutions which travel forward in $\alpha$. The field quanta
introduced in this way, however, do not coincide with the usual
Poincar\'e invariant definition of a particle. The expansion of
the field operator $\ophi(\alpha,\beta)$ in terms of the functions
$\phi_\lambda$ reads
\begin{equation}
\ophi(\alpha,\beta)=\int_{-\infty}^{\infty} \, d\lambda \, \left[
\ob_\lambda \, \phi_\lambda (\alpha, \beta)+\obdag_\lambda \,
\phi_\lambda^\ast (\alpha, \beta)\right] \, ,
\end{equation}
with the ($\alpha$-independent) \lq\lq Fourier coefficients\rq\rq\
$\ob_\lambda$ and $\obdag_\lambda$ being given by
\begin{equation}
\label{ccrl} \ob_\lambda = \left( \phi_\lambda, \ophi
\right)_\alpha \, , \qquad \obdag_\lambda = -\left(
\phi_\lambda^\ast, \ophi \right)_\alpha\, .
\end{equation}
Equations~(\ref{ccrl}) and the equal-$\alpha$ commutation
relations, Eqs.~(\ref{ccr}), imply the harmonic-oscillator
commutation relations
\begin{equation}
\left[\ob_\lambda, \obdag_{\lambda^\prime} \right] =
\delta(\lambda-\lambda^\prime)\, , \qquad \left[\ob_\lambda,
\ob_{\lambda^\prime} \right] =\left[\obdag_\lambda,
\obdag_{\lambda^\prime} \right] = 0\, .
\end{equation}
The operators $\obdag_\lambda$ and $\ob_\lambda$ can be
interpreted as creation and annihilation operators of field quanta
which are characterized by a real value $\lambda$. The physical
interpretation of $\lambda$ is that of an eigenvalue of $\hat K$,
the generator of Lorentz boosts. In the $\lambda$-basis the
operator $\hat{K}$, as calculated from the stress-energy tensor,
becomes diagonal~\cite{z05}. Its Fock-space representation is
\begin{equation}
\hat{K} = \int_{-\infty}^\infty d\lambda \, \lambda\,
\obdag_\lambda \ob_\lambda \, .
\end{equation}
This means in particular that $\vert \lambda \rangle =
\obdag_{\lambda} \vert 0 \rangle$, with $\vert 0 \rangle$ denoting
the vacuum state, is an eigenstate of $\hat{K}$, i.e. $\hat{K}
\vert \lambda \rangle = \lambda \vert \lambda \rangle$.

Unlike the boost generator, the dilatation generator
$\hat{D}(\alpha)$ is, in general, not diagonalized by the boost
eigenstates $\vert \lambda\rangle$. Even for the interaction-free
case its Fock-space representation has a complicated structure
(for brevity we have neglected the arguments $m e^\alpha$ of the
Hankel functions):
\begin{eqnarray}
\hat{D}(\alpha) &=& \frac{\pi}{8} m^2 e^{2
\alpha}\int_{-\infty}^{\infty} d\lambda \, \left\{ \left[ 2
H_{i\lambda}^{(2)} H_{i\lambda}^{(1)} - H_{i\lambda-1}^{(2)}
H_{i\lambda+1}^{(1)} - H_{i\lambda+1}^{(2)}
H_{i\lambda-1}^{(1)}\right]
\ob_\lambda \obdag_\lambda \right. \nonumber\\
&-& \left. 2 \left[ H_{i\lambda}^{(2)} H_{-i\lambda}^{(2)} +
H_{1+i\lambda}^{(2)} H_{1-i\lambda}^{(2)} \right] \ob_\lambda
\ob_\lambda + h.c. \right\} \, .
\end{eqnarray}
Single particle states are therefore not eigenstates of the
dilatation operator $\hat{D}(\alpha)$.

Our discussion has up till now been confined to the forward light
cone. As has been shown in Ref.~\cite{ds74b} this restriction can
be overcome by analytic continuation of $\phi_\lambda$ along
appropriately chosen (complex) paths for $\alpha$ and $\beta$. In
this way $\phi_\lambda$ is represented by 4 different functions,
each belonging to one of the 4 wedges of Minkowsi space-time.
Evolution of the (quantized) fields can then be considered from
surface to surface with the surfaces being hyperboloids in the
forward and backward light cone and cones for $x^2<0$,
respectively. This means that outside the light cone the
hyperbolic coordinates $\alpha$ and $\beta$ essentially exchange
their roles -- $\beta$ becomes the time parameter and $\alpha$
labels the position on the cone. As a consequence
$\hat{D}(\alpha)$ cannot be used outside the light cone, but
another generator for evolution in $\beta$ has to be introduced.
Altogether it does not seem to be very practical to study
evolution of quantum field theories in hyperbolic coordinates.
Exceptions
 are perhaps scale-invariant theories for which the mass
$m$ has to vanish. The limit $m\rightarrow 0$, however, does not
easily follow from the formulas given above, but has to be
considered separately~\cite{ds74b}.

Another problem with the kind of approach just outlined is
connected with the $\lambda$-representation. This representation
diagonalizes the boost generator $\hat{K}$, but it complicates
matters for the momentum operator. To see the reason we first
express the single-particle boost eigenstates $\vert \lambda
\rangle$ in terms of momentum eigenstates $\vert p \rangle$,
\begin{equation}
\vert \lambda \rangle = \int_{-\infty}^\infty \frac{dp}{2
\omega_p} \vert p \rangle \langle p \vert \lambda \rangle \, ,
\end{equation}
with $\omega_p = \sqrt{m^2+p^2}$ and
\begin{equation}
\label{eq:pltrans} \langle p \vert \lambda \rangle = (\phi_p\,
,\phi_\lambda)_\alpha =\frac{1}{\sqrt{\pi}}\left(
\frac{p+\omega_p}{m}\right)^{i\lambda} = \frac{1}{\sqrt{\pi}} e^{i
\lambda \chi}\, .
\end{equation}
For practical calculations it is often convenient to replace the
momentum $p$ by the variable $\chi$, which is defined via $p= m
\sinh \chi$ and $\omega_p = m \cosh \chi$. $\phi_p$ are usual
plane waves expressed in terms of hyperbolic coordinates
\begin{equation}
\phi_p(\alpha,\beta) = \frac{1}{\sqrt{2 \pi}} e^{-i(\omega_p
t(\alpha,\beta)-p x(\alpha,\beta))}=\frac{1}{\sqrt{2 \pi}} e^{- i
m \exp(\alpha)\cosh(\chi-\beta)}\, .
\end{equation}
With the help of Eq.~(\ref{eq:pltrans}) we can now see that the
spatial component of the momentum operator $\hat{P}^1$ shifts
$\lambda$ by an imaginary quantity
\begin{eqnarray}
\langle \lambda^\prime \vert \hat{P}^1 \vert \lambda \rangle &=&
\int_{-\infty}^\infty \frac{dp}{2\omega_p} \int_{-\infty}^\infty
\frac{dp^\prime}{2\omega_{p^\prime}} \, \langle \lambda^\prime
\vert p^\prime \rangle\langle p^\prime \vert \hat{P}^1 \vert p
\rangle \langle p \vert \lambda \rangle \nonumber
\\ & = & \int_{-\infty}^\infty \frac{dp}{2\omega_p} \int_{-\infty}^\infty
\frac{dp^\prime}{2\omega_{p^\prime}} \frac{1}{\sqrt{\pi}}\left(
\frac{p^\prime+\omega_{p^\prime}}{m}\right)^{-i\lambda^\prime}
\!\!\!\!\!\!\left(p 2\omega_p \delta(p-p^\prime)\right)
\frac{1}{\sqrt{\pi}}\left(
\frac{p+\omega_p}{m}\right)^{i\lambda}\nonumber\\ &=& \frac{1}{2
\pi} \int_{-\infty}^\infty \frac{dp}{\omega_p} p \left(
\frac{p+\omega_p}{m}\right)^{i(\lambda-\lambda^\prime)} =
\frac{m}{2\pi}\int_{-\infty}^\infty d\chi \sinh(\chi)\, e^{i\chi
(\lambda-\lambda^\prime)}\nonumber \\ & = &
\frac{m}{4\pi}\int_{-\infty}^\infty d\chi \left( e^{i\chi
(\lambda-\lambda^\prime-i)}-e^{i\chi
(\lambda-\lambda^\prime+i)}\right) \nonumber\\ &=&
\frac{m}{2}\left[ \delta(\lambda-\lambda^\prime-i)-
\delta(\lambda-\lambda^\prime+i)\right]\, .
\end{eqnarray}
For $\hat{P}^0$ the minus sign between the two delta functions has
to be replaced by a plus sign. The meaning of this result has been
clarified in Ref.~\cite{mr72}. Since $\hat{P}^\mu$ are unbounded
operators they are only defined on a subspace of the one-particle
Hilbert space $\mathcal{L}^2((-\infty,\infty); dp/(2 \omega_p))$.
By means of the transformation given in Eq.~(\ref{eq:pltrans})
this subspace goes over into a subspace of square integrable
functions in $\lambda$ that is characterized by the property that
its elements are analytic in a strip $\vert \mathrm{Im}\, \lambda
\vert < 1$. The matrix elements $\langle \lambda^\prime \vert
\hat{P}^\mu \vert \lambda \rangle$, $\mu =0,1$, therefore have to
be understood as distributions acting on square-integrable
functions in $\lambda$ which are analytic in the strip $\vert
\mathrm{Im}\, \lambda \vert < 1$. For such test functions the
action of $\hat{P}^\mu$ is well defined: $\hat{P}^\mu \vert
f\rangle = \vert f^\prime\rangle$ with $f^\prime(\lambda) =
\langle \lambda \vert f^\prime\rangle = m (f(\lambda + i)
+(-1)^\mu f(\lambda - i ))/2$. The situation seems to be similar
to the case where one studies unitary representations of
noncompact groups and tries to diagonalize the operators which do
not generate a compact subgroup~\cite{i69}. Altogether, the
definition of the translation generators as selfadjoint operators
acting on square integrable functions of $\lambda$ obviously needs
special care and is, at least, not completely straightforward.

The generalization of the quantization procedure sketched above to
(free) complex scalar and spin-1/2 fields in 3+1-dimensional
Minkowski space-time has been worked out in Refs.~\cite{s74,ds74b}
and in Ref.~\cite{grs74}, respectively. In 3+1 dimensions the
orthogonal set of basis functions $\phi_\lambda(\alpha,\beta)$
(for scalar fields) has to be replaced by another set of functions
which are labelled by 3 parameters and which now depend on 3
spatial coordinates (usually $\beta$, $\theta$, and $\varphi$).
References~\cite{s74,ds74b} take $\phi_{\lambda, z}$ with $z$
being an arbitrary complex number, whereas Ref.~\cite{grs74}
rather uses $\phi_{\lambda, l, m}$ with integers $l=0,1,2,\dots$
and $m=-l,-l+1,\dots,l-1,l$. In both cases $(1+\lambda^2)$, with
$0\leq \lambda < \infty$, have to be understood as eigenvalues of
the Casimir operator $\vec{K^2}-\vec{L}^2$ of the homogeneous
Lorentz group. Its unitary irreducible representations can
therefore be characterized by $\lambda$ and the parameters $z$ or
$(l,m)$ label different orthogonal sets of basis vectors of the
representation space. The problems due to hyperbolic coordinates
and with the $\lambda$-representation of the 4-momentum operator
are, of course, also present in the 3+1 dimensional case. These
problems can be avoided in Euclidean field theories. For the
Euclidean version of PFQFT the hyperbolic coordinates are replaced
by spherical coordinates, which can be defined globally.
Quantization is then done on a sphere and evolution is considered
into radial direction. Since boosts are described by 4-dimensional
rotations in the Euclidean theory, the boost eigenvalues $\lambda$
assume integer values. The $\lambda$-representation of the
4-momentum operator contains still non-diagonal terms, but
$\hat{P}^\mu$ shifts $\lambda$ only by $\pm 1$. A discussion of
the Euclidean formulation for massless spin-0 and spin-1/2 fields
in 2 and 4 dimensions can be found in Ref. ~\cite{f73}. The
corresponding massive cases in 2 dimensions are considered in
Ref.~\cite{ds74a}.

Nowadays the radial quantization procedure of Fubini et
al.~\cite{f73} is the standard method for quantization of
2-dimensional, conformally symmetric models for which the
dilatation generator $\hat{D}$ is a constant of motion~\cite{k95}.
But it does not seem to be very useful in applications to massive
theories if one is not primarily interested in the spectrum of the
dilatation generator, but rather in the spectrum of the mass
operator. For this purpose it is more convenient to have a simple
representation of the 4-momentum operator $\hat{P}^\mu$ and pay
less attention to the dilatation generator $\hat{D}$. Also, a
Lorentz-invariant formulation of scattering does not necessarily
depend on evolution generated by $\hat{D}$, but can as well be
achieved in terms of the 4-momentum operator. In the following we
will thus make use of the usual basis of momentum eigenstates and
investigate the evolution of the system that is generated by
$\hat{P}^\mu$.

\section{Quantization of free fields}
\label{quantization}
In this section we will show for free field theories that the
usual (momentum) Fock-space representation of the Poincar\'e
generators, that is well known from equal-time quantization,
follows also from quantization on the hyperboloid $x_\mu x^\mu =
\tau^2$. We will start with the case of a complex scalar field.

\subsection{Spin-0 fields}
\label{spin_zero}
The Lagrangian density for a free scalar field in 3+1 dimensional
Minkowski space-time is
\begin{equation}
\mathcal{L}_{\mathrm{free}}(x)=\left[ (\partial_\mu
\phi^\ast(x))(\partial^\mu \phi(x)) - m^2
\phi^\ast(x)\phi(x)\right]\, , \label{eq:Lscalarfree}
\end{equation}
where $x$ denotes the contravariant 4-vector
$(x^\mu)=(t,\vec{x})$. The equation of motion which follows from
$\mathcal{L}_{\mathrm{free}}(x)$ is the Klein-Gordon equation
\begin{equation}
\label{eq:kg4} \left( \partial_\mu \partial^\mu +
m^2\right)\phi(x)=0 \,.
\end{equation}
An important statement which can be made for arbitrary solutions
$\psi(x)$ and $\chi(x)$ of the Klein-Gordon equation, Eq.
~(\ref{eq:kg4}), is the following:

\noindent The scalar product
\begin{eqnarray}
\label{eq:scalarp} (\psi,\chi)_\sigma &:=& i \int_\sigma
d\sigma^\mu(x) \left[ \psi^\ast(x)\partial_\mu \chi(x) - \chi(x)
\partial_\mu \psi^\ast(x)\right]\nonumber\\
&=&i \int_\sigma d\sigma^\mu(x) \left[ \psi^\ast(x)
\stackrel{\leftrightarrow}{\partial}_\mu \chi(x)\right]\, ,
\end{eqnarray}
with $\sigma$ denoting a space-like hypersurface of Minkowski
space-time, does not depend on $\sigma$.

\noindent The general proof of this statement can be found in
Ref.~\cite{s61}. For a hyperplane of fixed time,
\begin{equation}
\sigma_t: \, x^0=t \quad \Longrightarrow \quad d\sigma^\mu = d^3x
\, g^{0 \mu} ,
\end{equation}
Eq.~(\ref{eq:scalarp}) represents nothing else than the well known
fact that the scalar product
\begin{equation}
(\psi,\chi)_{\sigma_t} = i \int_{\mathbb{R}^3}  d^3x \left[
\psi^\ast(t^\prime,\vec{x})
\stackrel{\leftrightarrow}{\partial}_{t^\prime}
\chi(t^\prime,\vec{x})\right]_{t^\prime=t}
\end{equation}
does not depend on $t$. Now we will demonstrate that the scalar
product $(\psi,\chi)_\sigma$ for
\begin{equation}
\label{eq:hyperboloid} \sigma_\tau: \, x_\mu x^\mu = \tau^2\quad
\Longrightarrow \quad d\sigma^\mu(x) = 2\,d^4x\, \delta(x\cdot x
-\tau^2) \theta(x^0) x^\mu\, ,
\end{equation}
is independent of the chosen hyperboloid, i.e. it does not depend
on $\tau$. Since every solution $\phi(x)$ of the Klein-Gordon
equation~(\ref{eq:kg4}) may be decomposed into plane waves
\begin{equation}
\phi(x)=\int_{\mathbb{R}^3}\frac{d^3p}{(2 \pi)^{3/2}\, 2
\omega_{\vec{p}}}\, \tilde{\phi}(\vec{p})\, e^{-i x\cdot p} \, ,
\end{equation}
with $p=(p^\mu)=(\omega_{\vec{p}}\, ,\vec{p})$ and
$\omega_p=\sqrt{m^2+\vec{p}^{\, 2}}$, the scalar
product~$(\psi,\chi)_\sigma$ can be written as
\begin{eqnarray}
(\psi,\chi)_{\sigma_\tau} &=& i \int_{\mathbb{R}^4} 2\, d^4x\,
\delta(x\cdot x -\tau^2) \theta(x^0) x^\mu \left[ \psi^\ast(x)
\stackrel{\leftrightarrow}{\partial}_\mu \chi(x)\right]
\nonumber\\ &=& \int_{\mathbb{R}^3}\frac{d^3p^\prime}{(2
\pi)^{3/2}\, 2 \omega_{\vec{p}^{\, \prime}}}\,
\int_{\mathbb{R}^3}\frac{d^3p}{(2 \pi)^{3/2}\, 2
\omega_{\vec{p}}}\, \tilde{\psi}^\ast(\vec{p}^{\, \prime})
\, \tilde{\chi}(\vec{p}) \nonumber\\
& & \times \int_{\mathbb{R}^4} 2\, d^4x\, \delta(x\cdot x -\tau^2)
\theta(x^0) x^\mu \, (p^\prime + p)_\mu\, e^{i x\cdot
(p^\prime-p)}\, .
\end{eqnarray}
The $x$-integral is now nothing else than a Lorentz invariant
distribution $W(P,Q)$, with $P=(p^\prime + p)$ and $Q=(p^\prime -
p)$, which can be easily calculated in an appropriate frame. For
its calculation and its properties we refer to App.~\ref{WPQ}. By
means of Eq.~(\ref{eq:Wfinal}) we end up with
\begin{eqnarray}
\label{eq:scalarprod2} (\psi,\chi)_{\sigma_\tau}  &=&
\int_{\mathbb{R}^3}\frac{d^3p^\prime}{(2 \pi)^{3/2}\, 2
\omega_{\vec{p}^{\, \prime}}}\, \int_{\mathbb{R}^3}\frac{d^3p}{(2
\pi)^{3/2}\, 2 \omega_{\vec{p}}}\, \tilde{\psi}^\ast(\vec{p}^{\,
\prime}) \,
\tilde{\chi}(\vec{p})\, W(p^\prime+p\, ,p^\prime-p)\nonumber\\
&=& \int_{\mathbb{R}^3}\frac{d^3p}{2 \omega_{\vec{p}}}\,
\tilde{\psi}^\ast(\vec{p}) \, \tilde{\chi}(\vec{p})\, ,
\end{eqnarray}
which does not depend on $\tau$. Eq.~(\ref{eq:scalarprod2})
implies, in particular, that usual plane waves (
$\tilde{\psi}(\vec{p})\propto \delta^3(\vec{p}-\vec{p}^{\,
\prime})$ and $\tilde{\chi}(\vec{p})\propto
\delta^3(\vec{p}-\vec{p}^{\, \prime\prime})$) are orthogonal on
the hyperboloid $\sigma_\tau$.

In order to quantize the scalar field theory on the hyperboloid
$\sigma_\tau$ we demand the following (Lorentz-invariant)
quantization conditions for the field operator $\hat{\phi}(x)$:
\begin{equation}
\label{eq:ccr1} x^\mu \left[\hat{\phi}(y),\partial_\mu
\hat{\phi}^\dag(x) \right]_{x^2=y^2=\tau^2} = \,i \, x^0
\delta^3(\vec{y}-\vec{x})\, ,
\end{equation}
\begin{equation}
\label{eq:ccr2} \left[\hat{\phi}(y),\hat{\phi}(x)
\right]_{x^2=y^2=\tau^2} =
\left[\hat{\phi}^\dag(y),\hat{\phi}^\dag(x)
\right]_{x^2=y^2=\tau^2} = 0\, .
\end{equation}
These quantization conditions generalize the 1+1 dimensional case,
Eqs.~(\ref{ccr}), without making reference to a particular choice
of a time parameter. Noting that $\partial/\partial\tau = (x\cdot
x)^{-1/2} x^\mu\partial_\mu$ and going over to Cartesian
coordinates, these quantization conditions are seen to agree with
those of Ref.~\cite{grs74}. They can as well  be considered as an
unintegrated version of Schwinger's covariant quantization
conditions adapted to the hyperboloid~\cite{s48}. The next step is
to expand the field operator $\hat{\phi}(x)$ in terms of a
complete set of solutions of the Klein-Gordon equation, which are
orthogonal under the invariant scalar product defined by
Eqs.~(\ref{eq:scalarp}) and (\ref{eq:hyperboloid}). As we have
seen before, we are allowed to take usual plane waves and write
($p^0=\omega_{\vec{p}}=\sqrt{m^2+\vec{p}^{\, 2}}$)
\begin{equation}
\label{expansion1} \hat{\phi}(x)=\int_{\mathbb{R}^3}\frac{d^3p}{(2
\pi)^{3/2}\, 2 \omega_{\vec{p}}}\, \left[\,\hat{a}(\vec{p})\,
e^{-i x\cdot p} + \hat{b}^\dag(\vec{p})\, e^{i x\cdot
p}\,\right]\, ,
\end{equation}
with the $\tau$ independent \lq\lq Fourier coefficients\rq\rq\
given by
\begin{eqnarray}
\hat{a}(\vec{p}) &=& \left( \phi_{\vec{p}},
\hat{\phi}\right)_{\sigma_\tau} \, , \quad \hat{a}^\dag(\vec{p}) =
- \left( \phi_{\vec{p}}^\ast,
\hat{\phi}^\dag\right)_{\sigma_\tau}\, ,
\nonumber\\
\hat{b}(\vec{p}) &=& \left( \phi_{\vec{p}},
\hat{\phi}^\dag\right)_{\sigma_\tau} \, , \quad
\hat{b}^\dag(\vec{p}) = - \left( \phi_{\vec{p}}^\ast,
\hat{\phi}\right)_{\sigma_\tau}\, ,
\end{eqnarray}
and $\phi_{\vec{p}}(x)=\exp(-i\, x\cdot p)/(2\pi)^{3/2}$. These
relations and the field commutators, Eqs.~(\ref{eq:ccr1}) and
(\ref{eq:ccr2}), imply the harmonic-oscillator commutation
relations
\begin{equation}
\label{eq:ccr3} \left[\hat{a}(\vec{p}),\hat{a}^\dag(\vec{p}^{\,
\prime}) \right] = \left[\hat{b}(\vec{p}),\hat{b}^\dag(\vec{p}^{\,
\prime}) \right] = 2\, \omega_{\vec{p}} \,
\delta^3(\vec{p}-\vec{p}^{\, \prime})\, .
\end{equation}
All the other commutators vanish. In deriving these commutation
relations one encounters two integrations over space-time
hyperboloids. One of these integrations is done by means of the
delta function resulting from the field commutators, the remaining
one is of the form $W(P,Q)$ or $W(Q,P)$ and can thus also been
done easily (cf. App.~\ref{WPQ}). Due to their commutation
relations the operators $\hat{a}(\vec{p})$, $\hat{b}(\vec{p})$,
$\hat{a}^\dag(\vec{p})$, and $\hat{b}^\dag(\vec{p})$ may be
interpreted as lowering or raising operators which annihilate or
create field quanta characterized by a continuous label $\vec{p}$.
As a result of the $U(1)$-symmetry of our scalar theory we expect
the existence of a conserved charge operator
\begin{equation}
\hat{Q} = i \int_{\mathbb{R}^4} 2\, d^4x\, \delta(x\cdot x
-\tau^2) \theta(x^0) x^\mu : \hat{\phi}^\dag(x)
\stackrel{\leftrightarrow}{\partial}_\mu \hat{\phi}(x): \, ,
\end{equation}
where \lq\lq$:\dots:$\rq\rq\ denotes usual normal ordering.
Inserting the field expansion, Eq.~(\ref{expansion1}), we get
again $x$-integrals of the form $W(P,Q)$, or $W(Q,P)$ which can be
used to do one of the $p$-integrations. Finally one ends up with
the usual form for the charge operator
\begin{equation}
\hat{Q} = \int_{\mathbb{R}^3}\frac{d^3p}{2
\omega_{\vec{p}}}\,\left[\, \hat{a}^\dag(\vec{p}) \hat{a}(\vec{p})
- \hat{b}^\dag(\vec{p}) \hat{b}(\vec{p}) \, \right] \, .
\end{equation}
This suggests that $\hat{a}^\dag(\vec{p})$ can be considered as a
 creation operator for particles of charge $+1$ and
$\hat{b}^\dag(\vec{p})$ as a creation operator of antiparticles
with charge $-1$.

Next we will express the generators of space-time translations in
terms of annihilation and creation operators. The starting point
is the energy-momentum tensor (for brevity we neglect the argument
of $\ophi(x)$):
\begin{eqnarray}
\hat{T}^{\mu\nu}(x)&=:&\partial^{\mu}\ophi^{\dag}\partial^{\nu}
\ophi+\partial^{\nu}\ophi^{\dag}\partial^{\mu}\ophi-
g^{\mu\nu}\mathcal{L}\left(\ophi,\ophi^\dag,
\partial_\alpha\ophi,\partial_\alpha\ophi^\dag\right):.
\nonumber\\ & & \label{eq:eptensor}
\end{eqnarray}
For later purposes we note that this expression is also valid for
interacting theories as long as the interaction terms do not
contain derivatives of the fields. The 4-momentum operator, which
generates translations of the field operator in the sense that
$\partial^\mu\ophi(x) = i\, [\hat{P}^\mu,\ophi(x)]$, is then given
by
\begin{equation}
\hat{P}^\mu = \int_{\sigma_\tau} d\sigma_\nu\,
 \hat{T}^{\nu\mu}(x)
= \int_{\mathbb{R}^4} 2\, d^4x\, \delta (x^2-\tau^2 )\theta (x^0
)x_{\nu}\, \hat{T}^{\nu\mu}(x)\, .\label{eq:pep}
\end{equation}
Taking the Lagrangian density, Eq.~(\ref{eq:Lscalarfree}), and
inserting the field expansion, Eq.~(\ref{expansion1}), into
$\hat{T}^{\mu\nu}(x)$ yields
\begin{eqnarray}
\hat{P}^{\mu}_{\mathrm{free}}&=&\frac{1}{\left(2\pi\right)^3}\int_{\mathbb{R}^4}2\,
d^4x \, \delta(x^2-\tau^2)\theta(x^0)\, \int_{\mathbb{R}^3}
\frac{d^3p}{2\omega_{\vec{p}}}\, \int_{\mathbb{R}^3}
\frac{d^3p^\prime}{2\omega_{\vec{p}^{\,\prime}}}\nonumber\\
& &\left\{ \left[ {p^{\prime}}^{\mu} x\cdot p+p^{\prime}\cdot x
p^{\mu}-x^{\mu}\left( p\cdot p^\prime-m^2\right) \right]\right.
\nonumber\\ & &\times \left. \left( e^{i\left(
p^\prime-p\right)\cdot x }
\oa^{\dag}\left(\vec{p}^{\,\prime}\right)\oa\left(\vec{p}\right)+e^{-i\left(
p^\prime-p\right)\cdot x }
\ob^{\dag}\left(\vec{p}\right)\ob\left(\vec{p}^{\,\prime}\right)\right)\right.
\nonumber\\
 &-&\left.\left[ p'^{\mu} x\cdot p+p'\cdot x
p^{\mu}-x^{\mu}\left( p\cdot p'+m^2\right) \right]  \right. \nonumber\\
& &\times \left. \left( e^{i\left( p^\prime+p\right)\cdot x }
\oa^{\dag}\left(\vec{p}^{\,\prime}\right)\ob^{\dag}\left(\vec{p}\right)+
e^{-i\left(p^{\prime}+p\right)\cdot x }
\oa\left(\vec{p}\right)\ob\left(\vec{p}^{\,\prime}\right)\right)
\right\} \, .\label{eq:Pmu}
\end{eqnarray}
Introducing new momenta $P=p+p^\prime$ and $Q=p-p^\prime$ we may
rewrite the square brackets as $[P^\mu x\cdot P - Q^\mu x\cdot Q+
x^\mu Q\cdot Q]/2$ and $[P^\mu x\cdot P - Q^\mu x\cdot Q- x^\mu
P\cdot P]/2$, respectively. Interchanging integrations,
Eq.~(\ref{eq:Pmu}) becomes
\begin{eqnarray}
\hat{P}^{\mu}_{\mathrm{free}}&=&\frac{1}{2 \left(2\pi\right)^3}
\int_{\mathbb{R}^3} \frac{d^3p}{2\omega_{\vec{p}}}\,
\int_{\mathbb{R}^3}
\frac{d^3p^\prime}{2\omega_{\vec{p}^{\,\prime}}}\nonumber\\
&\times& \left[ P^\mu\, W(P,Q) - Q^\mu Q^\nu \, W_\nu(P,Q) +
Q\cdot Q \, W^\mu(P,Q) \right]\,  \oadag(\vec{p}^{\,\prime})
\oa(\vec{p})\nonumber\\&+& \left[ P^\mu\, W(P,-Q) - Q^\mu Q^\nu \,
W_\nu(P,-Q) + Q\cdot Q \, W^\mu(P,-Q) \right]\, \obdag(\vec{p})
\ob(\vec{p}^{\,\prime})\nonumber\\&-&\left[ P^\mu P^\nu\,
W_\nu(Q,P) - Q^\mu W(Q,P) - P\cdot P \, W^\mu(Q,P) \right]\,
\oadag(\vec{p}^{\,\prime}) \obdag(\vec{p})\nonumber\\&-&\left[
P^\mu P^\nu\, W_\nu(Q,-P) - Q^\mu \, W(Q,-P) - P\cdot P \,
W^\mu(Q,-P) \right]\, \oa(\vec{p})
\ob(\vec{p}^{\,\prime})\,,\nonumber\\ & &
\end{eqnarray}
with the $x$-integrals denoted as in App.~\ref{WPQ}. With the
results of App.~\ref{WPQ} it is easily seen that only the $P^\mu
W(P,Q)$ terms survive and the remaining terms cancel each other.
With the explicit expression for $W(P,Q)$, as given in
Eq.~(\ref{eq:Wfinal}), we finally recover the usual Fock-space
representation of the 4-momentum operator
\begin{eqnarray}
\hat{P}^{\mu}_{\mathrm{free}}&=&\frac{1}{2 \left(2\pi\right)^3}
\int_{\mathbb{R}^3} \frac{d^3p}{2\omega_{\vec{p}}}\,
\int_{\mathbb{R}^3}
\frac{d^3p^\prime}{2\omega_{\vec{p}^{\,\prime}}} P^\mu\left\{
W(P,Q)\, \oadag(\vec{p}^{\,\prime}) \oa(\vec{p}) + W(P,-Q)\,
\obdag(\vec{p}) \ob(\vec{p}^{\,\prime})\right\}\nonumber\\
&=& \frac{1}{2}\int_{\mathbb{R}^3}
\frac{d^3p}{2\omega_{\vec{p}}}\, \int_{\mathbb{R}^3}
\frac{d^3p^\prime}{2\omega_{\vec{p}^{\,\prime}}} P^\mu P^0
\delta^3(\vec{Q})\left\{
 \oadag(\vec{p}^{\,\prime}) \oa(\vec{p}) +
\obdag(\vec{p}) \ob(\vec{p}^{\,\prime})\right\}\nonumber\\ &=&
\int_{\mathbb{R}^3} \frac{d^3p}{2\omega_{\vec{p}}}\, p^\mu\left\{
\oadag(\vec{p}) \oa(\vec{p}) + \obdag(\vec{p})
\ob(\vec{p})\right\}\, .
\end{eqnarray}
From this form of the 4-momentum operator and the the commutation
relations, Eqs.~(\ref{eq:ccr1}), we conclude that the field quanta
created by $\oa^\dag(\vec{p})$ and $\ob^\dag(\vec{p})$ are
eigenstates of the 4-momentum operator with eigenvalues $p^\mu$.
The corresponding calculation for the boost and rotation
generators is somewhat more tedious, but leads also to the well
known result. This proves the equivalence of equal-time
quantization and quantization on the hyperboloid $x_\mu
x^\mu=\tau^2$ for the case of free scalar fields.

\subsection{Spin-1/2 fields}
\label{spin_half}
In the case of a free spin-1/2 field one can proceed in an
analogous way. The Lagrangian density
\begin{equation}
\mathcal{L}_{\mathrm{free}}(x)=  \bar{\psi}(x)\left(i \gamma^\mu
\partial_\mu
 - m \right) \psi(x)
\end{equation}
leads to the Dirac equation
\begin{equation}
\left(i\gamma^\mu\partial_\mu - m \right) \psi(x)=0\, .
\label{eq:dirac}
\end{equation}
For arbitrary solutions $\psi(x)$ and $\phi(x)$ of the Dirac
equation, Eq.~(\ref{eq:dirac}), the invariant scalar product on
the hyperboloid reads~\cite{s61}
\begin{eqnarray}
\left( \psi, \phi \right)_{\sigma_\tau} &=& \int_{\sigma_\tau}
d\sigma^\mu(x)\, \left[\bar{\psi}(x)\gamma_\mu
\phi(x)\right]\nonumber\\ &=& \int_{\mathbb{R}^4}2 \, d^4x\,
\delta(x^2-\tau^2)\theta(x^0)x^\mu\, \left[\bar{\psi}(x)\gamma_\mu
\phi(x)\right]\, .\label{eq:scalarprod3}
\end{eqnarray}
A set of appropriately normalized solutions, which are orthogonal
with respect to this scalar product, is given by
\begin{eqnarray}
\psi^{\left(+\right)}_{\lambda,\vec{p}}\left(x\right)=\frac{1}
{\left(2\pi\right)^{\frac32}}e^{-ip\cdot
x}u_{\lambda}\left(\vec{p}\right),\quad
\psi^{\left(-\right)}_{\lambda,\vec{p}}\left(x\right)=\frac{1}
{\left(2\pi\right)^{\frac32}}e^{ip\cdot
x}v_{\lambda}\left(\vec{p}\right)\,, \quad
\lambda=\pm\frac{1}{2}\, ,\nonumber\\ \label{pw2}
\end{eqnarray}
with
\begin{eqnarray}
u_{\lambda}\left(\vec{p}\right)=\sqrt{\omega_{\vec{p}}+m}\left(%
\begin{array}{c}
  \chi_\lambda \\
  \frac{\vec{\sigma}\cdot\vec{p}}{\omega_{\vec{p}}+m}\chi_\lambda \\
\end{array}
\right)\, , \quad
v_{\lambda}\left(\vec{p}\right)=-\sqrt{\omega_{\vec{p}}+m}\left(%
\begin{array}{c}
  \frac{\vec{\sigma}\cdot\vec{p}}{\omega_{\vec{p}}+m}\, \hat{\epsilon}
  \chi_\lambda \\ \hat{\epsilon} \chi_\lambda \\
\end{array}
\right)\, ,\nonumber\\
\end{eqnarray}
and
\begin{equation}
\chi_\lambda = \frac{1}{2}\left(\begin{array}{c}
  (1+2\lambda) \\
  (1-2\lambda)\\
\end{array}\right)\, , \quad \hat{\epsilon}=
\left(\begin{array}{rcr}
  0 && 1 \\
  -1 && 0 \\
\end{array}\right)\, .
\end{equation}
The 4-spinors are normalized such that
\begin{equation}
\bar{u}_{\lambda}\left(\vec{p}\right)\gamma^\mu u_{\lambda^\prime}
\left(\vec{p}\right)=
\bar{v}_{\lambda}\left(\vec{p}\right)\gamma^\mu v_{\lambda^\prime}
\left(\vec{p}\right)=2 p^\mu \delta_{\lambda\lambda^\prime}\,
.\label{eq:spinnorm}
\end{equation}

Quantization of the spin-1/2 field $\hat{\psi}(x)$ on the
hyperboloid $\sigma_\tau$ may be accomplished by demanding
(Lorentz-invariant) anticommutation relations of the form:
\begin{equation}
x^{\mu}\left\{
\hat{\psi}_{a}\left(y\right),[\,\hat{\!\bar{\psi}}\left(x\right)
\gamma_{\mu}]_{b}\right\}_{x^2=y^2=\tau^2}= x^0 \, \delta_{a b }\,
\delta^3\left(\vec{x}-\vec{y}\right) \, ,\label{eq:acr1}
\end{equation}
\begin{equation}
\left\{\hat{\psi}_{a}\left(y\right),\hat{\psi}_{b}
\left(x\right)\right\}_{x^2=y^2=\tau^2}=
\{\,\hat{\!\bar{\psi}}_a\left(y\right),
\,\hat{\!\bar{\psi}}_b\left(x\right)\}_{x^2=y^2=\tau^2}=0\,
.\label{eq:acr2}
\end{equation}
The subscripts \lq\lq$a$\rq\rq and \lq\lq$b$\rq\rq\ label the
components of the 4-spinors $\hat{\psi}$ and $\hat{\bar{\!
\psi}}$. As in the scalar case these quantization conditions are
consistent with those given in Refs.~\cite{s48,grs74}. Creation
and annihilation operators of field quanta with a particular
momentum $\vec{p}$ are introduced by expanding the field operator
$\hat{\psi}$ in terms of the plane waves given in Eq.~(\ref{pw2}):
\begin{equation}
\hat{\psi}\left(x\right)=\frac{1}{\left(2\pi\right)^{3/2}}
\sum_{\lambda=\pm 1/2}\, \int_{\mathbb{R}^3}
\frac{d^3p}{2\omega_{\vec{p}}}\,
\left(\hat{c}_{\lambda}\left(\vec{p}\right) u_{\lambda}
\left(\vec{p}\right)e^{-ip\cdot x}+ \hat{d}_{\lambda}^{\dag}
\left(\vec{p}\right) v_{\lambda}\left(\vec{p}\right)e^{ip\cdot x}
\right)\, .\label{eq:fourierpsi}
\end{equation}
The creation and annihilation operators are recovered by means of
the invariant scalar product, Eq.~(\ref{eq:scalarprod3}),
\begin{eqnarray}
\hat{c}_{\lambda}\left(\vec{p}\right)&=&\left(\psi_{\lambda,
\vec{p}}^{(+)} ,\hat{\psi}\right)_{\sigma_\tau}\, ,\quad
\hat{c}_{\lambda}^\dag
\left(\vec{p}\right)=\left(\hat{\psi},\psi^{(+)}_{\lambda,\vec{p}}
\right)_{\sigma_\tau},\nonumber\\
\hat{d}_{\lambda}\left(\vec{p}\right)&=&
\left(\hat{\psi},\psi_{\lambda,\vec{p}}^{(-)}
\right)_{\sigma_\tau},\quad
\hat{d}_{\lambda}^{\dag}\left(\vec{p}\right)=
\left({\psi}^{(-)}_{\lambda,\vec{p}}
,\hat{\psi}\right)_{\sigma_\tau}\, .\label{eq:cdop}
\end{eqnarray}
These relations are verified with the help of Eqs.~(\ref{eq:Wmu1})
-- (\ref{eq:Wmu2}), the spinor normalization condition,
Eq.~(\ref{eq:spinnorm}), and the fact that
$\bar{u}_{\lambda}(\vec{p})\gamma^\mu (p-p')_\mu
u_{\lambda^\prime} (\vec{p}^{\,
\prime})=\bar{u}_{\lambda}(\vec{p})\gamma^\mu (p+p')_\mu
v_{\lambda^\prime} (\vec{p}^{\, \prime})=0$. The anticommutation
relations for creation and annihilation operators follow from
Eqs.~(\ref{eq:cdop}) and the anticommutation relations for the
field operators, Eqs.~(\ref{eq:acr1}) and (\ref{eq:acr2}):
\begin{equation}
\label{eq:acr3}
\left\{\hat{c}_\lambda(\vec{p}),\hat{c}^\dag_{\lambda^\prime}(\vec{p}^{\,
\prime}) \right\} =
\left\{\hat{d}_\lambda(\vec{p}),\hat{d}_{\lambda^\prime}^\dag(\vec{p}^{\,
\prime}) \right\} = 2\, \omega_{\vec{p}} \,\, \delta_{\lambda
\lambda^{\prime}}\, \delta^3(\vec{p}-\vec{p}^{\, \prime})\, .
\end{equation}
All the other anticommutators vanish. For illustration we will in
the following prove the anticommutation relation of
$\hat{c}_\lambda(\vec{p})$ with
$\hat{c}^\dag_{\lambda^\prime}(\vec{p}^{\, \prime})$:
\begin{eqnarray}
\left\lbrace \hat{c}_{\lambda}(\vec{p}),
\hat{c}^{\dag}_{\lambda'}(\vec{p}^{\, \prime})\right\rbrace&=&
\left\{ \left(\psi_{\lambda, \vec{p}}^{(+)}
,\hat{\psi}\right)_{\sigma_\tau},
\left(\hat{\psi},\psi^{(+)}_{\lambda',\vec{p}^{\, \prime}}
\right)_{\sigma_\tau}\right\}\nonumber\\
&=& \int_{\mathbb{R}^4}2\, d^4x\, \delta (x^2-\tau^2)\,
\theta(x^0) \int_{\mathbb{R}^4}2\, d^4x'\, \delta (x'^2-\tau^2 )\,
\theta(x'^0 )\nonumber\\&& \times x^{\mu}x'^{\nu}
\left\lbrace\left[\bar{\psi}_{\lambda,\vec{p}}^{(+)}
 (x)\gamma_{\mu}\right]_{a}
\left[\hat{\psi} (x)\right]_a,\left[\hat{\!\bar{\psi}}
 (x')\gamma_{\nu}\right]_{b}\left[\psi_{\lambda',\vec{p}^{\, \prime}}^{(+)}
 (x')\right]_b \right\rbrace \nonumber\\&=& \frac{4}{(2 \pi)^3}
\int_{\mathbb{R}^4} d^4x\, \delta (x^2-\tau^2)\, \theta(x^0)
\int_{\mathbb{R}^4} d^4x'\, \delta (x'^2-\tau^2 )\, \theta(x'^0
)\nonumber\\&& \times  e^{ip\cdot x}e^{-ip'\cdot x'}x^{\mu}
\left[\bar{u}_{\lambda}(\vec{p}) \gamma_{\mu}\right]_{a}\left[
u_{\lambda'} (\vec{p}^{\, \prime})\right]_b \underbrace{x'^{\nu}
\left\lbrace\left[\hat{\psi}(x)\right]_a,
\left[\hat{\!\bar{\psi}}(x')\gamma_{\nu}\right]
_{b}\right\rbrace}_{=x^0\delta_{a b}\delta^3
\left(\vec{x}-\vec{x}^{\, \prime}\right)}\nonumber\\&=&
\frac{1}{\left(2\pi\right)^3}\left[\bar{u}_{\lambda}(\vec{p})\gamma_{\mu}
u_{\lambda'}(\vec{p}^{\, \prime})\right] \int_{\mathbb{R}^4}d^4x
\,2\, \delta (x^2-\tau^2)\,\theta(x^0) x^{\mu} e^{-i(p'-p)\cdot
x}\nonumber\\
&=& \frac{1}{\left(2\pi\right)^3}
\left[\bar{u}_{\lambda}(\vec{p})\gamma_{\mu}
u_{\lambda'}(\vec{p}^{\, \prime})\right]
W^\mu(p'+p,p'-p)\nonumber\\
&=& \underbrace{\left[\bar{u}_{\lambda}(\vec{p})\gamma_{\mu}
u_{\lambda'}(\vec{p})\right]}_{=2\, p_\mu \, \delta_{\lambda
\lambda'}}\, \frac{2 p^\mu}{4 p \cdot p}\, 2 \omega_{\vec{p}}\,
\delta^3(\vec{p}-\vec{p^{\,\prime}})\nonumber\\
&=& 2\, \omega_{\vec{p}} \, \delta_{\lambda
\lambda'}\,\delta^3(\vec{p}-\vec{p^{\,\prime}})\, .
\end{eqnarray}
Here we have used $\bar{u}_{\lambda}(\vec{p})\gamma^\mu (p-p')_\mu
u_{\lambda^\prime} (\vec{p}^{\, \prime})=0$, as well as
Eqs.~(\ref{eq:Wfinal}) and (\ref{eq:Wmu3}).

As in the scalar case we have a conserved charge operator
\begin{equation}
\hat{Q} = \int_{\mathbb{R}^4} 2\, d^4x\, \delta(x\cdot x -\tau^2)
\theta(x^0) x^\mu : \hat{\!\bar{\psi}} \gamma_\mu \hat{\psi}(x):
\, .\label{eq:chargef}
\end{equation}
After insertion of the field expansion, Eq.~(\ref{eq:fourierpsi}),
into Eq.(\ref{eq:chargef}), one of the 3-dimensional momentum
integrations can be performed by means of the delta function
coming from the integration over the hyperboloid. The resulting
Fock-space representation of the charge operator
\begin{equation}
\hat{Q} = \sum_{\lambda=\pm 1/2}\,\int_{\mathbb{R}^3}\frac{d^3p}{2
\omega_{\vec{p}}}\,\left[\, \hat{c}^\dag_\lambda(\vec{p})
\hat{c}_\lambda(\vec{p}) - \hat{d}^\dag_\lambda(\vec{p})
\hat{d}_\lambda(\vec{p}) \, \right]
\end{equation}
suggests that $\hat{c}^\dag(\vec{p})$ be considered as creation
operators of particles with charge $+1$ and
$\hat{d}^\dag(\vec{p})$ as creation operators of antiparticles
with charge $-1$.

An appropriate version of the energy momentum tensor for the free
spin-1/2 field is:
\begin{equation}
\hat{T}^{\mu\nu}_{\mathrm{free}}(x):=\frac{i}{2}\,
:\hat{\!\bar{\psi}}(x)\gamma^\mu
\stackrel{\leftrightarrow}{\partial}\!\!\!\phantom{i}^{\nu}
\hat{\psi}(x):\, .
\end{equation}
With this expression for $\hat{T}^{\mu\nu}(x)$ the Fock-space
representation of the 4-momentum operator becomes
(cf.~Eq.~(\ref{eq:pep})):
\begin{eqnarray}
\hat{P}^{\mu}_{\mathrm{free}}&=&\frac{i}{\left(2\pi\right)^3}\sum_{\lambda,\lambda'=\pm
1/2}\, \int_{\mathbb{R}^4} d^4x \, \delta(x^2-\tau^2)\,
\theta(x^0)
 \, \int_{\mathbb{R}^3} \frac{d^3p}{2\omega_{\vec{p}}}\,
\int_{\mathbb{R}^3}
\frac{d^3p^\prime}{2\omega_{\vec{p}^{\,\prime}}}\nonumber\\
& &\times :\left(\hat{c}^\dag_{\lambda'}(\vec{p}^{\,\prime})
\bar{u}_{\lambda} (\vec{p}^{\,\prime})e^{ip'\cdot x}+
\hat{d}_{\lambda} (\vec{p}^{\,\prime})
\bar{v}_{\lambda}(\vec{p}^{\,\prime})e^{-ip'\cdot x}
\right) x_\nu  \gamma^\nu \nonumber\\
& &\times
\stackrel{\leftrightarrow}{\partial}\!\!\!\phantom{i}^{\mu}
 \left(\hat{c}_{\lambda}(\vec{p})
u_{\lambda} (\vec{p})e^{-ip\cdot x}+ \hat{d}_{\lambda}^{\dag}
(\vec{p})
v_{\lambda}(\vec{p})e^{ip\cdot x} \right): \nonumber\\
&=& \frac{1}{\left(2\pi\right)^3}\sum_{\lambda,\lambda'=\pm 1/2}\,
\int_{\mathbb{R}^4} d^4x \, \delta(x^2-\tau^2)\, \theta(x^0)
 \, \int_{\mathbb{R}^3} \frac{d^3p}{2\omega_{\vec{p}}}\,
\int_{\mathbb{R}^3}
\frac{d^3p^\prime}{2\omega_{\vec{p}^{\,\prime}}}\nonumber\\
& &\times : \left\lbrace (p+p')^\mu \left(e^{ix\cdot (p'-p)}
\bar{u}_{\lambda'}(\vec{p}^{\, \prime}) x_\nu \gamma^{\nu}
u_{\lambda}(\vec{p})\, \hat{c}_{\lambda'}^{\dag} (\vec{p}^{\,
\prime})\hat{c}_{\lambda}(\vec{p})\right.\right. \nonumber\\& &
\hspace{2.4cm} \left.\left. -e^{-ix\cdot
(p'-p)}\bar{v}_{\lambda'}(\vec{p}^{\, \prime}) x_\nu \gamma^{\nu}
v_{\lambda}(\vec{p})\, \hat{d}_{\lambda'} (\vec{p}^{\,
\prime})\hat{d}_{\lambda}^{\dag}(\vec{p}) \right)\right.\nonumber
\\
&&\left.\hspace{0.6cm}+(p-p')^\mu \left(e^{-ix\cdot(p'+p)}
\bar{v}_{\lambda'}(\vec{p}^{\, \prime})x_\nu
\gamma^{\nu}u_{\lambda} (\vec{p})\, \hat{d}_{\lambda'}(\vec{p}^{\,
\prime})\hat{c}_{\lambda}(\vec{p}) \right.\right. \nonumber\\ &
&\left.\left.\hspace{2.4cm} -e^{ix\cdot (p'+p)}\bar{u}_{\lambda'}
(\vec{p}^{\, \prime})x_\nu \gamma^{\nu}v_{\lambda}(\vec{p}) \,
\hat{c}_{\lambda'}^{\dag} (\vec{p}^{\,
\prime})\hat{d}_{\lambda}^{\dag}
(\vec{p})\right)\right\rbrace :\nonumber\\
&=& \frac{1}{2 (2\pi )^3}\sum_{\lambda,\lambda'=\pm 1/2}\,
\int_{\mathbb{R}^3} \frac{d^3p}{2\omega_{\vec{p}}}\,
\int_{\mathbb{R}^3}
\frac{d^3p^\prime}{2\omega_{\vec{p}^{\,\prime}}}\nonumber\\
& &\times : \left\lbrace P^\mu W_\nu(P,-Q)\,
\bar{u}_{\lambda'}(\vec{p}^{\, \prime})\gamma^{\nu}
u_{\lambda}(\vec{p})\,\hat{c}_{\lambda'}^{\dag} (\vec{p}^{\,
\prime})\hat{c}_{\lambda}(\vec{p})\right. \nonumber\\& &
\hspace{0.6cm} \left. - P^\mu W_\nu(P,Q)\,
\bar{v}_{\lambda'}(\vec{p}^{\, \prime}
)\gamma^{\nu}v_{\lambda}(\vec{p})\,\hat{d}_{\lambda'} (\vec{p}^{\,
\prime})\hat{d}_{\lambda}^{\dag}(\vec{p}) \right.\nonumber
\\
&&\hspace{0.6cm}+ Q^\mu W_\nu(Q,P)\,
\bar{v}_{\lambda'}(\vec{p}^{\, \prime})\gamma^{\nu}u_{\lambda}
(\vec{p})\,\hat{d}_{\lambda'}(\vec{p}^{\, \prime})
\hat{c}_{\lambda}(\vec{p}) \nonumber\\
& &\left.\hspace{0.6cm} - Q^\mu W_\nu(Q,-P)\, \bar{u}_{\lambda'}
(\vec{p}^{\, \prime})\gamma^{\nu}v_{\lambda}(\vec{p}
)\,\hat{c}_{\lambda'}^{\dag}(\vec{p}^{\,
\prime})\hat{d}_{\lambda}^{\dag}
(\vec{p})\right\rbrace :\nonumber\\
&=& \frac{1}{2}\sum_{\lambda,\lambda'=\pm 1/2}\,
\int_{\mathbb{R}^3} \frac{d^3p}{2\omega_{\vec{p}}}\,
\int_{\mathbb{R}^3}
\frac{d^3p^\prime}{2\omega_{\vec{p}^{\,\prime}}}\,
2\omega_{\vec{p}^{\,\prime}}\, \delta^3(\vec{p}-\vec{p}^{\,
\prime})\, \frac{P^\mu P_\nu}{P\cdot P}
\nonumber\\
& &\times  \left\lbrace \bar{u}_{\lambda'}(\vec{p}^{\,
\prime})\gamma^{\nu}
u_{\lambda}(\vec{p})\,\hat{c}_{\lambda'}^{\dag} (\vec{p}^{\,
\prime})\hat{c}_{\lambda}(\vec{p}) +
\bar{v}_{\lambda'}(\vec{p}^{\,
\prime})\gamma^{\nu}v_{\lambda}(\vec{p})\,\hat{d}_{\lambda}^{\dag}(\vec{p})
\hat{d}_{\lambda'} (\vec{p}^{\, \prime}) \right\rbrace\nonumber\\
&=& \sum_{\lambda=\pm 1/2}\, \int_{\mathbb{R}^3}
\frac{d^3p}{2\omega_{\vec{p}}}\,p^\mu\,
\left\{\hat{c}_{\lambda}^{\dag}
(\vec{p})\hat{c}_{\lambda}(\vec{p})+\hat{d}_{\lambda}^{\dag}
(\vec{p})\hat{d}_{\lambda}(\vec{p})\right\}\, .
\end{eqnarray}
Here we have again used the the fact that
$\bar{u}_{\lambda'}(\vec{p}^{\, \prime})Q_\nu \gamma^\nu
u_{\lambda} (\vec{p})=\bar{v}_{\lambda'}(\vec{p}^{\, \prime})
Q_\nu \gamma^\nu v_{\lambda}(\vec{p})$ $= \bar{u}_{\lambda'}
(\vec{p}^{\, \prime})P_\nu \gamma^\nu v_{\lambda}(\vec{p})=
\bar{v}_{\lambda'} (\vec{p}^{\, \prime})P_\nu \gamma^\nu
u_{\lambda}(\vec{p})=0$, the spinor normalization condition,
Eq.~(\ref{eq:spinnorm}), and the properties of $W_\nu(P,Q)$ and
$W_\nu(Q,P)$, respectively (with $P=p+p'$, $Q=p-p'$). Thus we have
proved in the spin-1/2 case that the Fock-space representation of
the 4-momentum operator takes on its well known form. With some
more effort this can also be verified for rotation and boost
generators which finally establishes the equivalence of equal-time
quantization and quantization on the hyperboloid $x\cdot x =
\tau^2$ for free spin-1/2 fields.

\section{Interacting fields and scattering}
\label{scattering}
After having shown that quantization on the forward hyperboloid
and equal-time quantization provide the same Fock-space
representation of the Poincar\'e generators for free fields (if
the same set of basis states is taken), we will now investigate
the effect of including an interaction term into the Lagrangian
density, i.e. $\mathcal{L}(x)=\mathcal{L}_{\mathrm{free}}(x) +
\mathcal{L}_{\mathrm{int}}(x)$.
As long as $\mathcal{L}_{\mathrm{int}}(x)$ does
not contain derivatives of the fields, we infer immediately from
Eqs.~(\ref{eq:eptensor}) and (\ref{eq:pep}) that the interacting
part of the 4-momentum operator is given by
\begin{eqnarray}
\hat{P}^\mu_{\mathrm{int}} &=& - \int_{\sigma_\tau} d\sigma_\nu\,
 g^{\nu\mu}\, :\hat\mathcal{L}_{\mathrm{int}}(x):\,\,
= - \int_{\mathbb{R}^4} 2\, d^4x\, \delta (x^2-\tau^2 )\theta (x^0
)\, x^{\mu} :\hat\mathcal{L}_{\mathrm{int}}(x):\, ,\nonumber\\ & &
\label{eq:Pint}
\end{eqnarray}
where $\hat\mathcal{L}_{\mathrm{int}}(x)$ is short for
$\mathcal{L}_{\mathrm{int}}(\hat{\phi}(x),\partial_\mu
\hat{\phi}(x),\dots)$. This means that all components of the
4-momentum operator become interaction dependent when quantizing
on the hyperboloid $\sigma_\tau$. Let us next look at the
generators of spatial rotations and Lorentz boosts. If we combine
them to the antisymmetric tensor $\hat{M}^{\mu\nu}$ it is not
difficult to see that the interaction dependent part of this
tensor vanishes:
\begin{eqnarray}
\hat{M}^{\mu\nu}_{\mathrm{int}} &=& \int_{\sigma_\tau}
d\sigma_\rho\,
 \left[x^\mu \hat{T}^{\rho\nu}_{\mathrm{int}}  -
 x^\nu \hat{T}^{\rho\mu}_{\mathrm{int}} \right]\nonumber\\
&=& - \int_{\mathbb{R}^4} 2\, d^4x\, \delta (x^2-\tau^2 )\theta
(x^0 )\, x_{\rho}\left[ x^\mu g^{\rho\nu} - x^\nu
g^{\rho\mu}\right] :\hat\mathcal{L}_{\mathrm{int}}(x):\,\, = 0 \,
.
\end{eqnarray}
When quantizing on the hyperboloid $\sigma_\tau$ the rotation and
boost generators are thus not affected by interactions.

For the discussion of scattering in interacting quantum field
theories it is most convenient to use the interaction picture. In
the interaction picture the evolution of operators is the same as
for the free system whereas the interaction determines the
evolution of the states. If one considers the evolution of the
system generated by the 4-momentum operator, the interaction
picture in PFQFT can be cast into a nice covariant form, by
splitting the 4-momentum operator into a free and an interacting
part,
\begin{equation}
\hat{P}^\mu =
\hat{P}^\mu_{\mathrm{free}}+\hat{P}^\mu_{\mathrm{int}}\, ,
\end{equation}
where each term is obtained by quantizing on the forward
hyperboloid.
 Since all components of the 4-momentum operator are
interaction dependent it makes sense to adapt the interaction
picture in such a way that it covers evolution into arbitrary
space-time directions. Let $\hat{\mathcal{O}}$ be an operator and
$|\psi \rangle$ be a state specified on the quantization surface
$\sigma_\tau$. Evolution of the system into the $x_\mu$-direction
is then described by the set of equations
\begin{eqnarray}
\label{eq:intpict1} i \partial^\mu \hat\mathcal{O}(x)&=& \left[\,
\hat\mathcal{O}(x)\, , \hat{P}_{\mathrm{free}}^\mu  \right]\, ,
\quad \hat\mathcal{O}(x=0)=\hat\mathcal{O}\, ,\\ && \nonumber\\
i\partial^\mu \vert \psi(x) \rangle &=&
\hat{P}_{\mathrm{int}}^\mu(x) \vert \psi(x) \rangle\, , \quad
\vert \psi(x=0) \rangle = \vert \psi \rangle\, ,
\label{eq:intpict2}
\end{eqnarray}
where
\begin{equation}
\hat{P}_{\mathrm{int}}^\mu(x):= e^{i \hat{P}_{\mathrm{free}} \cdot
x}\, \hat{P}_{\mathrm{int}}^\mu \, e^{-i \hat{P}_{\mathrm{free}}
\cdot x}\, . \label{eq:P(t)}
\end{equation}
Equation~(\ref{eq:intpict2}) is formally solved by introducing an
evolution operator $\hat{U}(y,x)$ such that
\begin{equation}
\hat{U}(y,x)\vert \psi(x)\rangle = \vert \psi(y)\rangle\, .
\end{equation}
Applying $\partial^\mu$ to both sides of this equation we infer
with the help of Eq.~(\ref{eq:intpict2}) that $\hat{U}(y,x)$ has
to satisfy the differential equation
\begin{equation}
i\frac{\partial}{\partial y_\mu} \hat{U}(y,x) =
\hat{P}^\mu_{\mathrm{int}}(y) \hat{U}(y,x)\, , \quad
\mathrm{with}\quad \hat{U}(x,x)=\hat{1}\, .
\end{equation}
For further purposes it is more convenient to rewrite this
initial-value problem for $\hat{U}(x,x_0)$ as an integral
equation:
\begin{equation}
\hat{U}(y,x)=\hat{1}-i\int_{\mathcal{C}(x,\,y)}\!\!\!
dy_\mu^\prime\, \hat{P}^\mu_{\mathrm{int}}(y^\prime)\,
\hat{U}(y^\prime,x)\, .
\end{equation}
The integral runs along an arbitrary smooth path
$\mathcal{C}(x,\,y)$ joining $x$ with $y$. The formal solution of
this integral equation can be written as a path-ordered
exponential
\begin{equation}
\hat{U}(y,x)=\mathcal{P}\exp\left(
-i\int_{\mathcal{C}(x,\,y)}\!\!\! dy_\mu^\prime\,
\hat{P}^\mu_{\mathrm{int}}(y^\prime)\, \right)\, .
\label{eq:Ucov}\end{equation} This result may be used to define
the scattering operator:
\begin{equation}
\hat{S}=\lim_{x^2,\, y^2\, \rightarrow \, +\infty} \hat{U}(y,x)
\quad \mathrm{such~that}\quad x^0<0\, ,\, y^0>0\, ,
\label{eq:Scov}
\end{equation}
i.e. the limits are taken in such a
way that $y$ stays in the forward and $x$ in the backward light
cone, respectively. Since the path $\mathcal{C}(x,\,y)$ can be
chosen arbitrarily, we can take a straight line joining $x$ and
$y$. The path for the calculation of the scattering operator may
thus be parameterized as
\begin{equation}
y_\mu^\prime(s)=a_\mu + s \, n_\mu \label{eq:sdep}
\end{equation}
with the  timelike 4-vector $n$ given by
\begin{equation}
n = \lim_{x^2,\, y^2\, \rightarrow \, +\infty}
\frac{y-x}{\sqrt{(y-x)^2}}\, , \quad \mathrm{such\ that} \quad
n\cdot n=1\, ,
\end{equation}
and $a_\mu$ appropriately chosen. With this parameterization the
scattering operator becomes a simple $s$-ordered exponential
(which we indicate by $\mathcal{S}$ in front of the exponential)
\begin{equation}
\hat{S} = \mathcal{S}\exp\left( -i\int_{-\infty}^\infty ds\,
n_\mu\, \hat{P}^\mu_{\mathrm{int}}(y^\prime(s))\, \right)\, .
\label{eq:Sop}
\end{equation}

In order to check whether Eq.~(\ref{eq:Sop}) provides a sensible
definition of the scattering operator we first expand the
exponential up to leading order in the interaction:
\begin{equation}
\hat{S} = \hat{1}-i\int_{-\infty}^\infty ds\, n_\mu\,
\hat{P}^\mu_{\mathrm{int}}(y^\prime(s))\, \, + \dots
\end{equation}
By means of Eqs.~(\ref{eq:Pint}) and (\ref{eq:P(t)}) we get
\begin{eqnarray}
\hat{S} &=& \hat{1}+i\int_{-\infty}^\infty ds\, n_\mu\,
\int_{\mathbb{R}^4} 2\, d^4x\, \delta (x^2-\tau^2 )\theta (x^0 )\,
x^{\mu} :\hat\mathcal{L}_{\mathrm{int}}(x+y^\prime(s)): + \dots \nonumber\\
&=& \hat{1}+i\int_{-\infty}^\infty ds\, \int_{\mathbb{R}^3}
\frac{d^3x}{x^0} \, n\cdot x\,
:\hat\mathcal{L}_{\mathrm{int}}(x+a+s n): + \dots \, .
\end{eqnarray}
Since $n$ is a timelike vector it can be written as $n={\Lambda}_v
\tilde{n}$ with $\tilde{n}=(1,0,0,0)$ and ${\Lambda}_v$ an
appropriate boost. Further, taking into account that $d^3x/x^0$ is
a Lorentz invariant integration measure the integral over the
Lagragian density can be written as
\begin{equation}
\hat{S} = \hat{1}+i\int_{-\infty}^\infty ds\, \int_{\mathbb{R}^3}
d^3\tilde{x}
:\hat\mathcal{L}_{\mathrm{int}}\left({\Lambda}_v(\tilde{x}+\tilde{a}+s
\tilde{n})\right):+\dots \, ,
\end{equation}
with $\tilde{x}=\hat{\Lambda}^{-1}_v x$ and
$\tilde{a}=\hat{\Lambda}^{-1}_v a$. After a further change of
coordinates ($z=(\sqrt{\tau^2+\vec{\tilde{x}}^{\, 2}}+\tilde{a}^0
+ s, \vec{\tilde{x}}+\vec{\tilde{a}}\,)$, $d^4z = ds\,
d^3\tilde{x}$) we finally obtain
\begin{eqnarray}
\hat{S} = \hat{1}+i\int_{\mathbb{R}^4} d^4z\,
:\hat\mathcal{L}_{\mathrm{int}}\left({\Lambda}_v\,z\right):+\dots
 = \hat{1}+i\int_{\mathbb{R}^4} d^4z^\prime \,
:\hat\mathcal{L}_{\mathrm{int}}(z^\prime):+\dots\, ,\nonumber\\
\end{eqnarray}
with $z^\prime={\Lambda}_v\,z$. This result, however, is nothing
else than the familiar leading-order perturbative expression for
the scattering operator.

Beyond leading order in the interaction $s$-ordering comes into
play. The second-order contribution to the scattering operator is,
e.g., given by
\begin{eqnarray}
\hat{S}^{(2)} &=& (-i)^2 \int_{-\infty}^\infty ds_1\, n_\mu\,
\hat{P}^\mu_{\mathrm{int}}(y^\prime(s_1))\, \int_{-\infty}^{s_1}
ds_2\, n_\nu\,
\hat{P}^\nu_{\mathrm{int}}(y^\prime(s_2))\nonumber\\
&=& \frac{(-i)^2}{2!} \int_{-\infty}^\infty ds_1
\int_{-\infty}^{\infty} ds_2 \,  n_\mu  n_\nu \, \mathcal{S}
\left[\, \hat{P}^\mu_{\mathrm{int}}(y^\prime(s_1))\,
\hat{P}^\nu_{\mathrm{int}}(y^\prime(s_2))\, \right]\nonumber\\
&=& \frac{(-i)^2}{2!} \int_{-\infty}^\infty \! ds_1\!
\int_{-\infty}^{\infty} ds_2 \! \int_{\mathbb{R}^4}\! 2  d^4x_1
\delta (x_1^2-\tau^2 )\theta (x_1^0 )   \int_{\mathbb{R}^4} \! 2
d^4x_2  \delta (x_2^2-\tau^2)\theta (x_2^0)\nonumber\\
&&\times n_\mu x_1^{\mu}\, n_\nu x_2^\nu \left[
\theta(s_1-s_2):\hat\mathcal{L}_{\mathrm{int}}(x_1+a+s_1 n):
:\hat\mathcal{L}_{\mathrm{int}}(x_2+a+s_2 n): \right.\nonumber\\
\nonumber\\ &&\phantom{n_\mu x_1^{\mu}\, n_\nu
x_2^\nu}+\left.\theta(s_2-s_1):\hat\mathcal{L}_{\mathrm{int}}(x_2+a+s_2
n):
:\hat\mathcal{L}_{\mathrm{int}}(x_1+a+s_1 n):\right]\nonumber\\
&=&\frac{(-i)^2}{2!} \int_{-\infty}^\infty \! ds_1\!
\int_{-\infty}^{\infty} ds_2 \! \int_{\mathbb{R}^3}\!
d^3\tilde{x}_1 \int_{\mathbb{R}^3} \! d^3\tilde{x}_2\nonumber\\
&&\times\left[\theta(s_1-s_2)
:\hat\mathcal{L}_{\mathrm{int}}\left({\Lambda}_v(\tilde{x}_1+\tilde{a}+s_1
\tilde{n})\right)::\hat\mathcal{L}_{\mathrm{int}}
\left({\Lambda}_v(\tilde{x}_2+\tilde{a}+s_2
\tilde{n})\right):\right.\nonumber\\ & &\nonumber\\
&&\left.  +\theta(s_2-s_1)
:\hat\mathcal{L}_{\mathrm{int}}\left({\Lambda}_v(\tilde{x}_2+\tilde{a}+s_2
\tilde{n})\right)::\hat\mathcal{L}_{\mathrm{int}}
\left({\Lambda}_v(\tilde{x}_1+\tilde{a}+s_1
\tilde{n})\right):\right],\nonumber\\\label{eq:S2}
\end{eqnarray}
with $\tilde{n}$, ${\Lambda}_v$, $\tilde{a}$ defined as in the
leading-order case and $x_i={\Lambda}_v \tilde{x}_i$. With a
further change of coordinates
($z_i=(\sqrt{\tau^2+\vec{\tilde{x}}_i^{\, 2}}+\tilde{a}^0 + s_i,
\vec{\tilde{x}}_i+\vec{\tilde{a}}\,)$, $d^4z_i = ds\,
d^3\tilde{x}_i$, $i=1,2$) and the abbreviation
$d_{1,2}=\left(\sqrt{\tau^2+(\vec{z}_1-\vec{\tilde{a}})^2}-
\sqrt{\tau^2+(\vec{z}_2-\vec{\tilde{a}})^2}\right)$
Eq.~(\ref{eq:S2}) becomes
\begin{eqnarray}
\hat{S}^{(2)} &=& \frac{(-i)^2}{2!} \int_{\mathbb{R}^4} \! d^4 z_1
\int_{\mathbb{R}^4} \! d^4 z_2 \left[ \right.
\theta(z_1^0-z_2^0-d_{1,2})
:\hat\mathcal{L}_{\mathrm{int}}\left({\Lambda}_v z_1
\right)::\hat\mathcal{L}_{\mathrm{int}}
\left({\Lambda}_v z_2 \right):\nonumber\\
&& \qquad\qquad + \left. \theta(z_2^0-z_1^0+d_{1,2})
:\hat\mathcal{L}_{\mathrm{int}}\left({\Lambda}_v z_2
\right)::\hat\mathcal{L}_{\mathrm{int}} \left({\Lambda}_v z_1
\right): \right]\nonumber\\
&=& \frac{(-i)^2}{2!} \left\{\int_{\mathbb{R}^4} \! d^4 z_1
\int_{\mathbb{R}^4} \! d^4 z_2 \left[ \right. \theta(z_1^0-z_2^0)
:\hat\mathcal{L}_{\mathrm{int}}\left({\Lambda}_v z_1
\right)::\hat\mathcal{L}_{\mathrm{int}}
\left({\Lambda}_v z_2 \right):\right.\nonumber\\
&& \qquad\qquad + \left. \theta(z_2^0-z_1^0)
:\hat\mathcal{L}_{\mathrm{int}}\left({\Lambda}_v z_2
\right)::\hat\mathcal{L}_{\mathrm{int}} \left({\Lambda}_v z_1
\right): \right]\nonumber\\
&&\qquad + \int_{\mathbb{R}^4} \! d^4 z_1 \int_{\mathbb{R}^3} d^3
z_2 \!\! \int_{z_1^0}^{z_1^0-d_{1,2}} \!\! dz_2^0
:\hat\mathcal{L}_{\mathrm{int}}\left({\Lambda}_v z_1
\right)::\hat\mathcal{L}_{\mathrm{int}} \left({\Lambda}_v z_2
\right):\nonumber\\
&&\qquad + \left. \int_{\mathbb{R}^4} \! d^4 z_1
\int_{\mathbb{R}^3} d^3 z_2 \!\! \int^{z_1^0}_{z_1^0-d_{1,2}} \!\!
dz_2^0 :\hat\mathcal{L}_{\mathrm{int}}\left({\Lambda}_v z_2
\right)::\hat\mathcal{L}_{\mathrm{int}} \left({\Lambda}_v z_1
\right):\right\}\, .
\end{eqnarray}
If we now concentrate on the last two integrals we observe, that
due to the restriction on the $z_2^0$-integration $z_2$ and $z_1$
are always separated by a spacelike distance. Therefore
$:\hat\mathcal{L}_{\mathrm{int}}\left({\Lambda}_v z_1 \right):$
and $:\hat\mathcal{L}_{\mathrm{int}}\left({\Lambda}_v z_2
\right):$ can be interchanged and the two integrals cancel each
other. For the remaining integral we make a final change of
variables $z_i^\prime = \Lambda_v z_i$ and make use of
$\theta({z_i^0}^\prime-{z_j^0}^\prime)=\theta(z_i-z_j)$ to obtain
\begin{eqnarray}
\hat{S}^{(2)} &=&  \frac{(-i)^2}{2!} \int_{\mathbb{R}^4} \! d^4
z_1^\prime \int_{\mathbb{R}^4} \! d^4 z_2^\prime \left[
\theta({z_1^0}^\prime-{z_2^0}^\prime)
:\hat\mathcal{L}_{\mathrm{int}}\left(z_1^\prime
\right)::\hat\mathcal{L}_{\mathrm{int}}
\left(z_2^\prime \right):\right.\nonumber\\
&& \left. \qquad\qquad\qquad\qquad +
\theta({z_2^0}^\prime-{z_1^0}^\prime)
:\hat\mathcal{L}_{\mathrm{int}}\left({z_2}^\prime
\right)::\hat\mathcal{L}_{\mathrm{int}} \left({z_1}^\prime
\right): \right] \, .
\end{eqnarray}
But this is again nothing else than the second-order contribution
for the S-operator in usual time-ordered perturbation theory. The
way of reasoning just outlined for the two lowest orders can be
generalized to higher orders in the interaction so that
Eq.~(\ref{eq:Sop}) is indeed seen to be equivalent to the well
known (instant-form) representation of the S operator as the usual
time-ordered exponential~\cite{s61}. The equivalence of
Eq.~(\ref{eq:Sop}) with the usual perturbative expression for the
S-operator tells us also that total 4-momentum conservation is
guaranteed by our formulation of scattering, although 3-momentum
conservation at the vertices does, in general, not hold within
PFQFT. It remains to be seen whether any advantages  can be drawn
from the manifestly covariant representation of the S-operator, as
given in Eq.~(\ref{eq:Sop}) (or even more general in
Eqs.~(\ref{eq:Ucov}) and (\ref{eq:Scov})), to organize
perturbative calculations.

\section{Summary and Outlook}
\label{summary}
Point-form quantum field theory was first developed in the 1970's
by taking the forward hyperboloid $x_\mu x^\mu =\tau^2$ as the
quantization surface. All 4 components of the momentum operator
are then dynamic in the sense that they evolve the system away
from the quantization surface. The generators of Lorentz
transformations, on the other hand, are purely kinematic. It
seemed thus to be quite natural to choose a Fock-space basis which
is related to the generators of the Lorentz group (in contrast to
the more usual momentum basis used in instant- and front-form
quantum field theories). The advantage of this \lq\lq Lorentz
basis\rq\rq\ is that the quantum numbers which label the basis
states are conserved at interaction vertices, whereas 4-momentum
is not. In these earlier papers emphasis was put on studying the
evolution in $\tau$, i.e. the evolution generated by the
dilatation operator. However, as was realized very soon, and as is
discussed in Sec.~\ref{history}, the evolution in $\tau$ together
with the Lorentz basis leads to a number of conceptual
difficulties which stopped the further development of point-form
quantum field theory.

It is, nevertheless, still possible to carry out the
Schwinger-Tomonaga program for quantizing on a curved hypersurface
like the forward hyperboloid.  In this paper we have developed a
point-form quantum field theory in a momentum basis. In such a
basis overall energy {\bf and} 3-momentum are, in general, not
conserved in intermediate states. Neither is the free overall
4-velocity.  But overall 4-momentum conservation holds, of course,
between the (asymptotic) initial and final states, as our
perturbative analysis of the scattering operator has revealed.
This is to be contrasted with point-form relativistic quantum
mechanics for finite degree-of-freedom systems, where the overall
free 4-velocity is usually chosen to be conserved at each
interaction vertex~\cite{bt53}.  Then overall energy and momentum
conservation is achieved from overall mass conservation. Thus,
when interacting mass operators in point form relativistic quantum
mechanics are obtained from quantum field theoretic vertices, the
requirement of 4-velocity conservation must be added as an
explicit requirement~\cite{k03b}.

We have shown how to analyze free (spin 0 and 1/2) quantum fields,
where the inner product is given by integration over the forward
hyperboloid (see Sec.~\ref{quantization}).  With such an
integration surface canonical quantization can be formulated in a
manifestly Lorentz covariant way, without making reference to a
particular time parameter. When interactions arising from products
of local fields are generated, all of the interactions are in the
4-momentum operator, and Lorentz generators are kinematic. A
convenient way to express the fact that quantization of a local
field theory on the forward hyperboloid provides a representation
of the Poincar\'e algebra are the \lq\lq point-form\rq\rq\
equations,
\begin{eqnarray}
[\hat{P}^{\mu},\hat{P}^{\nu}]&=&0\\&&\nonumber\\ \hat{U}_{\Lambda}
\hat{P}^{\mu} \hat{U}_{\Lambda}^{-1}&=&(\Lambda^{-1})^{\mu}_{\nu}
\, \hat{P}^{\nu},
\end{eqnarray}
where $\hat{P}^{\mu}$ is the total four-momentum operator
(including all interactions) and $\hat{U}_{\Lambda}$ is the
unitary operator implementing the Lorentz transformations.
Adopting the Schr\"odinger picture (indicated by a subscript
\lq\lq S\rq\rq), these equations lead naturally to a covariant
Schr\"odinger equation,
\begin{equation}
i\partial^\mu |\psi(x)\rangle_\mathrm{S} =
\hat{P}^{\mu}|\psi(x)\rangle_\mathrm{S} \, ,
\end{equation}
where $|\psi(x=0)\rangle_\mathrm{S}$ is the state of the system
specified on the quantization surface. This covariant
Schr\"odinger equation can be used for the solution of the quantum
field theoretic bound-state problem. If the total four-momentum
operator is written as a free plus interacting part,
$\hat{P}^{\mu}=\hat{P}^{\mu}_\mathrm{free}+\hat{P}^{\mu}_\mathrm{int}$,
a generalized interaction representation is easily obtained, which
is the starting point for the covariant formulation of scattering
given in Sec.~4.

The nice feature that the operator formalism becomes manifestly
Lorentz covariant if fields are quantized on the forward
hyperboloid is not the only reason to study point-form quantum
field theory. Relativistic quantum mechanical models (with a
finite number of particles) often rely on field theoretical ideas.
Thus it is quite natural to take point-form quantum field theory
as a starting point for the construction of effective
interactions, currents, etc., which can be applied to point-form
quantum mechanics. A further motivation for developing a point
form quantum field theory is to analyze gauge theories. Because
Lorentz transformations are kinematic and the theory is manifestly
Lorentz covariant, gauge transformations and gauge invariance can
be naturally incorporated into the theory. Thus, a promising
application will, e.g., be to view quantum chromodynamics as a
point form quantum field theory and investigate the nature of
gauge fixing and other properties of non-Abelian gauge theories.

\appendix

\section{The distribution $W(P,Q)$}
\label{WPQ}
The distribution $W(P,Q)$ is used repeatedly to perform
integrations over the space-time hyperboloid $x^2=\tau^2$. This
appendix summarizes the properties of $W(P,Q)$ needed for the
derivations in the various sections. To begin with, $W(P,Q)$ is
defined as
\begin{equation}
W(P,Q):= 2 \int d^4x\, \delta(x\cdot x-\tau^2)\, \theta(x_0)\,
x\cdot P\, e^{-ix\cdot Q}\, . \label{eq:W}
\end{equation}
In the relevant cases we have in addition $P\cdot Q =0$, $P$
timelike and $Q$ spacelike. Since $P$ is timelike, it can be
written as
\begin{equation}
P=\Boost(v)\left( \begin{array}{c}
  M \\
  \vec{0}
\end{array}\right) \quad \mathrm{with} \quad M^2= P_\mu P^\mu \, .
\end{equation}
$\Boost(v)$ is a rotationless canonical boost with velocity
$v=P/M$. Its explicit form is~\cite{k98}
\begin{equation}
\Boost(v)=\left(
\begin{array}{ccc}
  v^0& &\vec{v}^{\, T}\\
  \vec{v}& &\bold{1}+\frac{v^0-1}{\vec{v}^2}\vec
{v} \vec{v}^{\, T}\\
\end{array}
\right) = \left(
\begin{array}{ccc}
  P^0/M & &\vec{P}^{\, T}/M\\
  \vec{P}/M& &\bold{1}+\frac{P^0/M-1}{\vec{P}^2}\vec
{P} \vec{P}^{\, T}\\
\end{array}
\right)\, .
\end{equation}
Since $Q$ is orthogonal to $P$ we have
\begin{equation}
Q=\left( \begin{array}{c}
  Q^0 \\
  \vec{Q}
\end{array}\right) = \Boost(v)\left( \begin{array}{c}
  0 \\
  \vec{q}
\end{array}\right) \, .
\label{eq:Q}
\end{equation}
Inverting Eq.~(\ref{eq:Q}) gives\\
\begin{equation}
\vec{q}=\vec{Q}-\frac{\vec{v}\cdot\vec{Q}}{v_0+v_0^2}\, \vec{v}=
\bold{N}\, \vec{Q} \, ,
\end{equation}
where we have used the orthogonality of $P$ and $Q$ to express
$Q^0$ as $Q^0=\vec{v}\cdot \vec{Q}/v^0$. $\bold{N}$ is a $3\times
3$-matrix with elements $N_{ij}=\delta_{ij}-v_iv_j/(v_0+v_0^2)$
and determinant $\mathrm{det}(\bold{N})=1/v_0$. By definition
$W(P,Q)$ is Lorentz invariant (cf. Eq.~(\ref{eq:W})) so that it
can be easily calculated  in the frame where $P$ has vanishing
spacelike components:
\begin{eqnarray}
W(P,Q)&=& 2 \int d^4x\,
\frac{\delta(x^0-\sqrt{\tau^2+\vec{x}^2})}{2 x^0}\, \, x^0 M\,
e^{i\vec{x}\cdot \vec{q}}=\int d^3x\,
 M\,
e^{i\vec{x}\cdot \vec{q}}\nonumber\\ &=& (2 \pi)^3 M
\delta^3(\vec{q})\, .
\end{eqnarray}
Going back to the original frame we finally get
\begin{eqnarray}
W(P,Q)&=&(2 \pi)^3 M \delta^3(\vec{q})=(2 \pi)^3 M
\delta^3(\bold{N}\, \vec{Q})= (2 \pi)^3
\frac{M}{\mathrm{det}(\bold{N})} \delta^3(\vec{Q})\nonumber\\ &=&
(2 \pi)^3 M v_0\delta^3(\vec{Q})= (2 \pi)^3
P_0\delta^3(\vec{Q})\,,\quad P\cdot P>0\,\, \mathrm{and}\,\,
P\cdot Q=0 \, . \nonumber\\ & & \label{eq:Wfinal}
\end{eqnarray}

If $P$ and $Q$ are interchanged  $W(Q,P)$ becomes zero. Proceeding
in  the same way as before one has
\begin{eqnarray}
W(Q,P)&=& 2 \int d^4x\,
\frac{\delta(x^0-\sqrt{\tau^2+\vec{x}^2})}{2 x^0}\,
\vec{x}\cdot\vec{q}\, e^{-i x^0 M}\nonumber\\&=&\int d^3x\,
\frac{\vec{x}\cdot\vec{q}}{\sqrt{\tau^2+\vec{x}^2}}\, e^{-i
\sqrt{\tau^2+\vec{x}^2} M} = 0\,,\quad P\cdot P>0\,\,
\mathrm{and}\,\, P\cdot Q=0 \, , \nonumber\\ & &
\end{eqnarray}
since the integrand is odd in $\vec{x}$.

It is also useful to introduce a Lorentz vector
\begin{equation}
W_\mu(P,Q):= \frac{P_\mu}{P\cdot P}\, W(P,Q)+\frac{Q_\mu}{Q\cdot
Q}\, W(Q,Q)\, ,\label{eq:Wmu3}
\end{equation}
with $P\cdot P>0$, $P\cdot Q=0$, and $W(.,.)$ defined according to
Eq.~(\ref{eq:W}). Further, if Eq.(A.1) is differentiated with
respect to $P^{\mu}$, one encounters the distribution
\begin{equation}
W_\mu^{\tau}(Q) := 2 \int d^4x\, \delta(x\cdot x-\tau^2)\, \theta(x_0)\,
x_\mu\, e^{-ix\cdot Q}\, . \label{eq:Wmu1}
\end{equation}
It is then straightforward to show that
$W_{\mu}^{\tau}=W_{\mu}(P,Q)$. We introduce 2 additional spacelike
4-vectors $R$ and $S$ which, together with $P$ and $Q$, form an
orthogonal basis of Minkowski space. If one now represents $x_\mu$
in terms of these basis vectors, the right-hand side of
Eq.~(\ref{eq:Wmu1}) becomes
\begin{eqnarray}
\frac{P_\mu}{P\cdot P}\, W(P,Q)+\frac{Q_\mu}{Q\cdot Q}\, W(Q,Q)
+\frac{R_\mu}{R\cdot R}\, W(R,Q) +\frac{S_\mu}{S\cdot S}\,
W(S,Q)\, .\nonumber\\
\end{eqnarray}
When calculating $W(R,Q)$ and $W(S,Q)$ we can choose our
coordinate system for the integration variables such that the
spatial coordinate axes coincide with $Q$, $R$, and $S$,
respectively. For this choice of coordinates it is immediately
obvious that the integrand of $W(R,Q)$ is odd in the $R$-direction
and thus $W(R,Q)=0$. An analogous reasoning holds for $W(S,Q)$,
which proves Eq.~(\ref{eq:Wmu1}). The specific form of the Lorentz
scalar $W(Q,Q)$ is not as simple as that of $W(P,Q)$, but it is
not required in our derivations.

By a similar reasoning we see that
\begin{eqnarray}
W_\mu(Q,P) := \frac{P_\mu}{P\cdot P}\, W(P,P) = 2 \int d^4x\,
\delta(x\cdot x-\tau^2)\, \theta(x_0)\, x_\mu\, e^{-ix\cdot P}\,
.\nonumber\\ \label{eq:Wmu2}
\end{eqnarray}
Actually $W_\mu(Q,P)$ does not depend on $Q$. We have only kept
$Q$ in its argument to better exhibit symmetry properties under
exchange of $P$ and $Q$ in intermediate steps of our calculations.

\medskip

\noindent {\bf Acknowledgement:} We would like to thank F. Coester
for stimulating discussions and the decisive hint how to evaluate
the scalar product on the forward hyperboloid.

\end{document}